\documentclass[fleqn,usenatbib]{mnras}
\pdfoutput=1
\usepackage{hyperref}
\usepackage{newtxtext,newtxmath}
\usepackage[T1]{fontenc}
\usepackage{ae,aecompl}

\usepackage{float}
\usepackage{graphicx}	
\usepackage{grffile}
\usepackage{amsmath}	
\usepackage{pdflscape}
\usepackage{array}
\newif\ifblackandwhite
\usepackage{multirow}
\usepackage{multicol}
\usepackage{longtable}
\usepackage[utf8]{inputenc}
\usepackage{siunitx}
\usepackage{color,soul}
\newcolumntype{L}[1]{>{\raggedright\let\newline\\\arraybackslash\hspace{0pt}}m{#1}}
\newcolumntype{C}[1]{>{\centering\let\newline\\\arraybackslash\hspace{0pt}}m{#1}}
\newcolumntype{R}[1]{>{\raggedleft\let\newline\\\arraybackslash\hspace{0pt}}m{#1}}

\newcommand{\nhunits}{\,cm$^{-2}$}                                              
\newcommand{\nh}{$N_\mathrm{H}$}                                                
                                   
\newcommand{\flucgs}{erg~cm$^{-2}$}                                             

\newcommand{\nustar}{{\em NuSTAR}}

\title[Repeating behaviour of FRB 121102]{ Repeating behaviour of FRB 121102: periodicity, waiting times and energy distribution}
\author[M. Cruces et al.]{
M. Cruces,$^{1}$\thanks{E-mail: mscruces@mpifr-bonn.mpg.de (MC)}
L. G. Spitler,$^{1}$
P. Scholz,$^{2}$
R. Lynch,$^{3}$
A. Seymour,$^{3}$ J. W. T.\newauthor
Hessels,$^{4,5}$,
C. Gouiffès,$^{6}$
G. H. Hilmarsson,$^{1}$
M. Kramer$^{1}$ and
S. Munjal$^{1}$
\\
$^{1}$Max-Planck Institute for Radio Astronomy, Auf dem Hügel 69, D-53121 Bonn, Germany\\
$^{2}$Dunlap Institute for Astronomy \& Astrophysics, University of Toronto, 50 St.~George Street, Toronto, ON M5S 3H4, Canada\\
$^{3}$ Green Bank Observatory, PO Box 2, WV 24944, Green Bank, USA\\
$^{4}$Anton Pannekoek Institute for Astronomy, University of Amsterdam, Science Park 904, 1098 XH, Amsterdam, The Netherlands\\
$^{5}$ASTRON, Netherlands Institute for Radio Astronomy, Oude
Hoogeveensedijk 4, 7991 PD Dwingeloo, The Netherlands\\
$^{6}$CEA/DRF/irfu/Département d’Astrophysique, AIM, Université Paris-Saclay, Université Paris Diderot, Sorbonne Paris Cité, CNRS,\\
F-91191 Gif-sur-Yvette,France\\
}
\date{Accepted XXX. Received YYY; in original form ZZZ}
\pubyear{2020}
\begin{document}
\label{firstpage}
\pagerange{\pageref{firstpage}--\pageref{lastpage}}
\maketitle
\begin{abstract}

Detections from the repeating fast radio burst FRB 121102 are clustered in time, noticeable even in the earliest repeat bursts. Recently, it was argued that the source activity is periodic, suggesting that the clustering reflected a not-yet-identified periodicity. We performed an extensive multi-wavelength campaign with the Effelsberg telescope, the Green Bank telescope and the Arecibo Observatory to shadow the Gran Telescope Canaria (optical), NuSTAR (X-ray) and INTEGRAL (gamma-ray). We detected 36 bursts with Effelsberg, one with a pulse width of 39\,ms, the widest burst ever detected from FRB 121102. With one burst detected during simultaneous NuSTAR observations, we place a 5-$\sigma$ upper limit of $5\times10^{47}$ erg on the 3--79\,keV energy of an X-ray burst counterpart. We tested the periodicity hypothesis using 165-hr of Effelsberg observations and find a periodicity of 161$\pm$5 days. We predict the source to be active from 2020-07-09 to 2020-10-14 and subsequently from 2020-12-17 to 2021-03-24. We compare the wait times between consecutive bursts within a single observation to Weibull and Poisson distributions. 
We conclude that the strong clustering was indeed a consequence of a periodic activity and show that if the few events with millisecond separation are excluded, the arrival times are Poisson distributed. We model the bursts' cumulative energy distribution with energies from ${\sim}10^{38}$-$10^{39}$ erg and find that it is well described by a power-law with slope of $\gamma=-1.1\pm 0.2$. We propose that a single power-law might be a poor descriptor of the data over many orders of magnitude. 
\end{abstract}


\section{Introduction}
Fast radio bursts (FRBs) are an observational phenomenon consisting of bright flashes of millisecond duration, detected so-far exclusively at radio frequencies, where detections have been made - as of now - at frequencies as low as 328 MHz \citep{Pilia2020,Chawla2020} and as high as 8 GHz \citep{Gajjar2018}. Although the majority of the sources are seen as one-off events, there are a couple of FRBs known to show repeated bursts at a consistent sky position. Example of this is FRB 121102 \citep{Spitler2016}, the first known repeating source, which was localized by VLA observations \citep{Chatterjee2017} and posteriorly its position pinpointed to milliarcsecond precision by VLBI \citep{Marcote2017}, and associated to a low-metallicity dwarf galaxy at redshift $z$=0.193 by the Gemini North observatory \citep{Tendulkar2017}.\\
Since the discovery of the Lorimer burst in archival pulsar data from the Parkes radio telescope \citep{Lorimer2007}, huge advances in the FRB field have been made during the last years with the discovery of over 100 FRBs \citep{FRBcat2016}\footnote{\label{frbcat}http://frbcat.org/}, localizations to host galaxies \citep{Chatterjee2017,Bannister2019,Chimeloc2020,Ravi2019, Macquart2020}, rotation measure (RM) and polarization measurements - revealing sometimes up to 100\% linearly polarized pulses with changing RM - probing the highly magnetic environment where the bursts originate \citep{Michilli2018}, and most recently the discovery of an active phase with a periodicity of 16 days in the repetition of FRB 180916.J0158+65 \citep{Chimeperiodicity2020}, and a potential 157 days periodicity for FRB 121102 \citep{PeriodicityFRB121102}.\\

Nonetheless, their astrophysical origin remains a mystery. Among the most popular models we find: mergers of double neutron star (DNS) systems \citep{DNSmerger,DNSmerger_repeat}, young magnetars \citep{magnetar2019}, giant pulses from pulsars \citep{giantpulses} and highly magnetic pulsar-asteroid interactions \citep{pulsar-asteriod} (see the FRB theory catalogue\footnote{\label{frbtheory}https://frbtheorycat.org} for more examples). While for some progenitor scenarios the detections are restricted to the radio frequencies regime, such as giant pulses from pulsars \citep{Cordes2016} and NS-WD mergers \citep{Liu2018}, other models predict counterparts at multiple wavelengths. An example of this is the young magnetar model, in which an additional X-ray afterglow and optical counterpart for its supernova remnant are expected \citep{magnetar2019}.\\

SGR J1935+2154 provides a particularly interesting magnetar-FRB link. This soft-gamma-ray repeater located in the Galaxy was associated with strong radio bursts \citep{CHIME2020_SGR}, followed by an X-ray counterpart \citep{Mereghetti2020}. For other scenarios such as the NS-NS merger scenario, in addition to optical emission from the kilonova and X-ray from an afterglow, a soft gamma-ray burst and gravitational wave counterparts are predicted \citep{DNSmerger_repeat}.\\

Multi-wavelength campaigns have the potential to constrain the aforementioned scenarios and to provide insights on the mechanisms at work. Yet, such campaigns are not plausible for most FRBs given their positional uncertainties of several arcminutes. As of now, few non-repeating FRBs have localization down to arcsecond precision \citep{askap2020,DSA2019}, however, their one-off nature makes multi-wavelength observations extremely challenging. On the contrary, repeating FRBs with precise localizations allows triggering observations at other wavelengths based on activity detected in the radio frequencies. Multi-wavelength follow-up from repeating FRBs, such as FRB 121102, have been key to our current understanding of FRBs. They have provided further evidence for the extragalactic origin of FRBs,  by ruling out the presence of an intervening HII region responsible for the dispersion measure (DM) excess of FRBs \citep{Scholz2016}, and have placed limits on the X-ray emission in the 0.5-10 keV to be less than $3\times 10^{-11}$ erg cm$^{-2}$ \citep{Scholz2017}.\\

To understand the progenitors of FRBs, observations spanning multiple epochs allow for a long-term periodicity study, which can be indicative of the presence of rotating binary systems, and how the detected bursts distribute in energy and waiting times provide clues on the nature of the source originating such bursts. The high-energy bursts of magnetars and the giant radio pulses from pulsars have shown to have energy distributions well modeled by a power-law. For magnetars it is found a slope $\gamma$ of -0.6 to -0.7 \citep{Gogus1999,Gogus2000}, while for giant pulses from the Crab pulsar, is observed $\gamma=-2.0$ \citep{Popov2007,Bera2019}.\\
\\
Motivated by these, we have performed an extended follow-up on FRB 121102 using the 100-m Effelsberg (EFF) radio telescope from September 2017 to June 2020. Some of the epochs are part of a multi-wavelength follow-up campaign to shadow higher energy telescopes such as \nustar, INTEGRAL, and the Gran Canaria Telescope (GTC), with radio telescopes such as Effelsberg, the 100-m Robert C. Byrd Green Bank Telescope (GBT), and the 305-m Arecibo Observatory (AO). We describe the observation setup and the algorithms employed for the data processing in Section~\ref{sec:Obs}. In Section~\ref{sec:bursts_prop} we report the bursts and their properties. We use the simultaneous observations with \nustar\ to place limits on the X-ray emission, and use the observed epochs with Effelsberg to test the potential 157-d periodicity reported by \citet{PeriodicityFRB121102}. Additionally, we test how the bursts are distributed within an active window, and as a whole study their cumulative energy distribution. We discuss the implications of the encountered periodicity, how the time interval between bursts distributes, and the power-law fit to the energy distribution in Section~\ref{sec:discussion}. We summarize and conclude in Section~\ref{sec:conclusion}.

\section{Observations and search}\label{sec:Obs}
\begin{figure*}
\begin{center}
	\includegraphics[trim={0 0cm 0 0cm},width=0.9\textwidth]{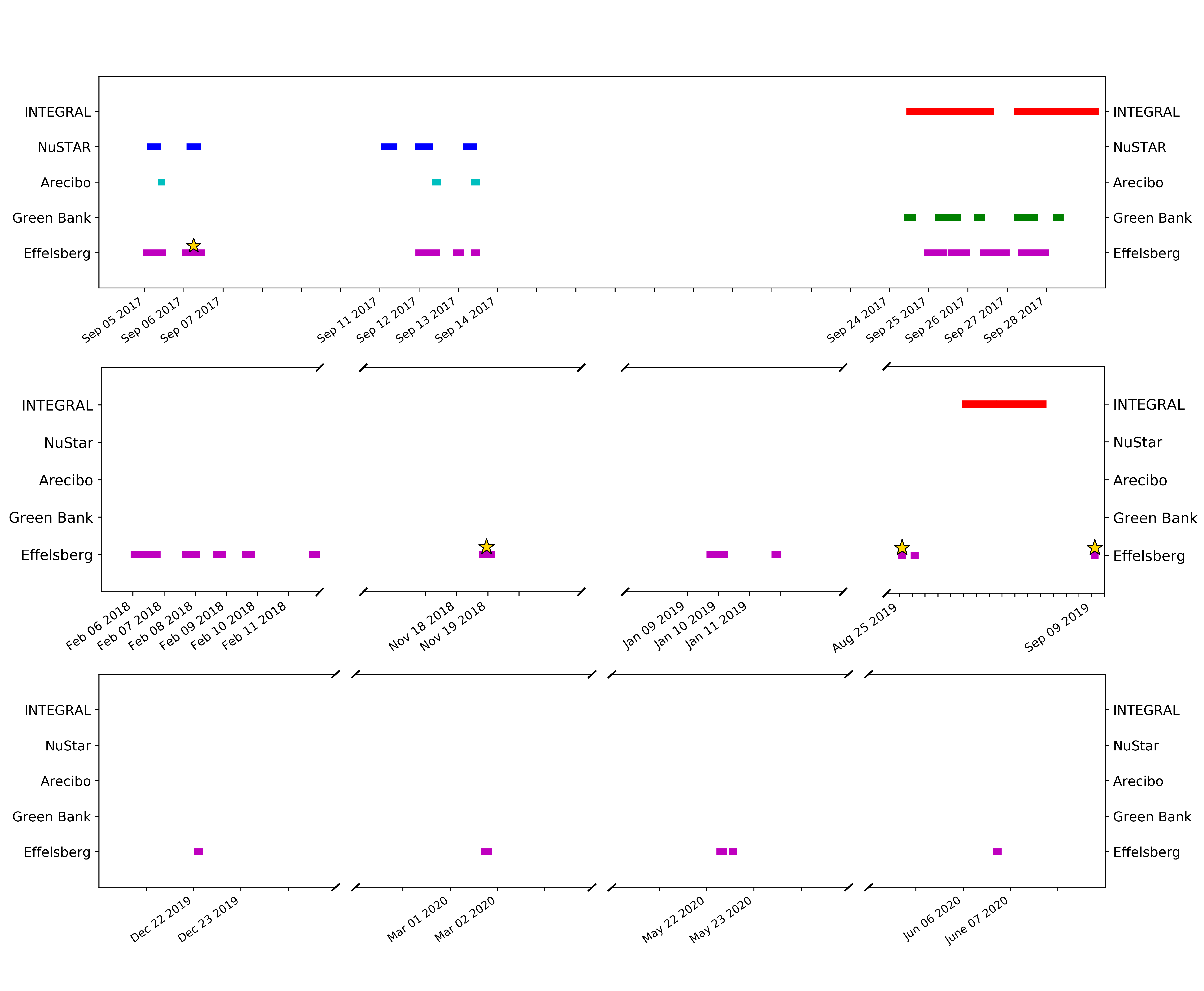}

\end{center}
\caption{Follow-up observations for FRB 121102 with Effelsberg (magenta), Green bank (green), Arecibo (cyan), NuSTAR (blue) and INTEGRAL (red). The yellow star marker indicates an epoch where at least one burst was detected.}
\label{fig:coverage}
\end{figure*}

Here we describe the radio observations simultaneous with three higher energy telescopes: \nustar, INTEGRAL, and GTC. The details of the radio observations are listed in Table~\ref{tab:obs} and the high energy observations in Table~\ref{tab:obs_highenergy}. The full coverage in frequency and time for FRB 121102 by the different telescopes is shown in Figure~\ref{fig:coverage}. It can be seen that in the case of the X-ray observations,  EFF and AO observed simultaneously with \nustar\ (Target ID: 80301307, PI: Scholz), and EFF and GBT observed simultaneously with INTEGRAL (project ID: 1420030, PI: GOUIFFES). For the optical observations with GTC, only EFF observed (project ID: 98-17, PI: Spitler). Such experiment had a similar approach to \citet{Hardy2017}, but with the GTC instead used for optical monitoring. The Effelsberg observations between 2018-02-05 to 2018-02-11 and 2019-01-09 to 2019-02-11 were scheduled together with the GTC to search for simultaneous optical bursts using HiPERCAM \citep{HiPERCAM2018}. No radio bursts were detected during the simultaneous optical observations, and the results from the analysis of the optical data alone are outside the scope of this paper.\\

FRB 121102 had an exposure time of 128 hours with Effelsberg, 26.25 hours with GBT, 3.7 hours with AO, 25.2 hours with \nustar\, and 240 hours with INTEGRAL. The only session with radio detections at the time of a simultaneous X-ray observation was on September 06, 2017, during a \nustar\, session (see section~\ref{sec:Nustar}). For some radio observations extended time coverage was performed. That is the case from the 24 to the 28 of September 2017 where EFF and GBT had almost full coverage, in some cases with simultaneous observing as well as one radio telescope taking over from the other. However, as seen in Figure~\ref{fig:coverage} no radio burst was detected in those observations.\\
We proceed hereon with the description of the observations and the data processing.

\begin{table*}
\caption{Description of the radio follow-up observations of FRB 121102. First column names the telescope (abbreviations: EFF=Effelsberg, GBT=Green Bank, AO=Arecibo observatory), second and third column list the starting time of the observation and its duration, and the fourth column shows the number of bursts detected. The observations for EFF were carried at a central frequency of 1.36 GHz, for AO at 4.5 GHz and for GBT at 2.0 GHz.}\label{tab:obs}
\begin{tabular}{ C{2cm} | C{2.6cm} C{1.6cm} |C{1.2cm} }
\hline
\hline
\multirow{1}{4em}{\textbf{Telescope}} & \multirow{2}{4em}{\textbf{Start} (UTC)} &\multirow{2}{4em}{\textbf{Duration }\centering{(s)}} &\multirow{1}{4em}{\textbf{Events}} \\
 & & & \\
\hline
EFF  & 2017-09-05 00:55:54  & 36000 & 0\\
AO & 2017-09-05 10:05 & 1009 & 0\\
EFF  & 2017-09-06 00:58:22 & 36000 & 7\\

EFF & 2017-09-11 23:49:16 & 39600 & 0\\
AO  & 2017-09-12 09:49 & 6118 & 0\\
EFF & 2017-09-12 22:53:54 & 8412 & 0\\
AO & 2017-09-13 09:46 & 6166 & 0\\
EFF & 2017-09-24 23:17:54 & 34854 & 0\\
GBT & 2017-09-24 10:45  & 11700 & 0\\

EFF & 2017-09-25 23:19:57 & 34992 & 0\\
GBT & 2017-09-25 06:00 & 7200 & 0\\
GBT & 2017-09-25 10:45 & 25200 & 0\\

EFF & 2017-09-26 23:28:52 & 32400 & 0\\ 
GBT & 2017-09-26 05:45  & 9900 & 0\\

EFF & 2017-09-27 23:26:49 & 23580 & 0\\
GBT & 2017-09-27 06:00  & 9000 & 0\\
GBT & 2017-09-27 10:45 & 22500 & 0\\

GBT & 2017-09-28 06:00 & 9000 & 0\\

EFF & 2018-02-05 18:36:47 & 22962 & 0\\

EFF & 2018-02-07 16:39:04 & 30018 & 0\\

EFF & 2018-02-08 16:46:25 & 16044 & 0\\

EFF & 2018-02-09 14:33:58 & 18000 & 0\\

EFF & 2018-02-11 18:20:54 & 13704 & 0\\

EFF & 2018-11-18 19:58:50 & 25200 & 24\\
EFF & 2019-01-09 17:29:53 & 39600 & 0\\
EFF & 2019-01-11 19:39:45 & 7590 & 0\\
EFF & 2019-08-25 04:10:13 & 6876 & 2 \\
EFF & 2019-08-26 03:38:19 & 4975 & 0 \\
EFF & 2019-09-09 04:59:27 & 3600 & 3 \\
EFF & 2019-12-22 01:42:15 & 5400 & 0 \\
EFF & 2020-03-01 17:28:46 & 7200 & 0 \\
EFF & 2020-05-22 06:37:34 & 7200 & 0 \\
EFF & 2020-05-22 13:03:29 & 7200 & 0 \\
\hline
\end{tabular}
\end{table*}

\begin{table*}
\begin{center}
\caption{Description  of  the  \nustar\, and INTEGRAL follow-up  observations  of  FRB 121102. The ID for \nustar\, corresponds to the observation ID, while for INTEGRAL the ID corresponds to the revolution number.}\label{tab:obs_highenergy}
\begin{tabular}{ccccc}
\hline
\hline
Telescope & ID  & Start time  & End time & Exposure time \\
          &     &  (UTC)        & (UTC)      & (s) \\
\hline
\nustar & 80301307002 & 2017-09-05 03:31:32 & 2017-09-05 13:39:33 & 15528 \\              
\nustar & 80301307004 & 2017-09-06 03:38:11 & 2017-09-06 13:54:41 & 17567 \\              
\nustar & 80301307006 & 2017-09-11 02:49:08 & 2017-09-11 14:30:23 & 21177 \\              
\nustar & 80301307008 & 2017-09-11 23:37:24 & 2017-09-12 11:48:38 & 20917 \\              
\nustar & 80301307010 & 2017-09-13 04:53:09 & 2017-09-13 14:37:12 & 15476 \\  
INTEGRAL & 1866 & 2017-09-24 12:22:33 & 2017-09-26 16:28:18 & 179092 \\       
INTEGRAL & 1867 & 2017-09-27 06:20:13 & 2017-09-29 07:19:30 & 171675 \\        
INTEGRAL & 2131 & 2019-08-30 04:07:16 & 2019-09-01 09:09:36 & 182321 \\
INTEGRAL & 2132 & 2019-09-01 19:28:59 & 2019-09-04 00:55:56 & 184376 \\
INTEGRAL & 2133 & 2019-09-04 11:16:37 & 2019-09-06 05:13:53 & 141549 \\
\hline
\end{tabular}
\end{center}
\end{table*}

\subsection{Effelsberg telescope}
We took data using the 7-beam feed array receiver with the  Pulsar Fast Fourier Transform Spectrometer \citep[PFFTS,][]{Barr2013}, and the high precision pulsar timing backend PSRIX \citep{Lazarus2016}. The PFFTS records data at a central frequency of 1.36 GHz with 300-MHz of bandwidth divided into 512 frequency channels and a time resolution of 54.613 $\mu s$. However, these data are not synchronized with a maser clock. To compensate, we recorded simultaneously the incoming data of the central beam with the backend used for pulsar timing, PSRIX, as it provides high precision time stamps. This backend's band is centered at 1.3589 GHz, with a bandwidth of 250 MHz divided into 256 channels and a time resolution of 51.2 $\mu s$. This allows us to obtain the precise time of arrivals (TOAs) of the detected bursts displayed in Table~\ref{tab:bursts}.
\\
While the central beam of the 7-beam receiver was pointing at FRB 121102 with right ascension $\alpha$ = 05:31:58.70 and declination $\delta$ = $+$33:08:52.5
00 \citep{Chatterjee2017}, all the remaining beams from the feed array were simultaneously recorded with the PFFTS for the purpose of radio frequency interference (RFI) mitigation.  
\label{sec:Eff}  
\subsubsection{Single pulse search}
We searched for single pulses in the time series from 500 pc cm$^{-3}$ up to 600 pc cm$^{-3}$, with steps of 1 pc cm$^{-3}$, using a pipeline based on the \textit{pulsar search software} PRESTO \citep{Ransom2011}. The timeseries were downsampled by a factor 16 to match the intrachannel dispersion delay at 1.510 GHz, corresponding to the top of the frequency band. Candidates down to a signal-to-noise (S/N) of 6 were explored leading to a total of 36 bursts detected in the PFFTS data (see Figures~\ref{fig:septemberbursts2017}, \ref{novemberbursts2018} and \ref{augustbursts2019}).\\
\\
To calculate the accurate topocentric TOAs, we run a similar single-pulse search on PSRIX data and cross-match with the bursts detected in the PFFTS. This is due to the lag between the recording of the backends of ${\sim}$15 seconds. Barycenter TOAs (t$_{\rm{bary}}$) are afterwards calculated as follows:
\setlength{\abovedisplayskip}{15pt} \setlength{\abovedisplayshortskip}{0pt}
\begin{equation}\label{eq:delay}
    \mathrm{t_{bary}} =  \mathrm{t_{topo}} - \Delta D/f^2 + \Delta_{R\sun}
\end{equation}
where t$_{\rm{topo}}$ is the topocentric time of arrival at the telescope, in this case Effelsberg. The second term is the delay caused by dispersion due to the interstellar medium (ISM), which depends on the observing frequency $f$, and $\Delta D=4.1488008\times 10^{3}$ MHz $^2$pc$^{-1}$ cm$^3  \times\,\rm{DM}$. The third and last term on equation (\ref{eq:delay}), $\Delta_R{\sun}$, is the Römer delay, which is the light-travel time between the telescope (for Effelberg longitude = \ang{6.882778}, latitude =\ang{ 50.52472}) and the solar system barycenter.

\subsubsection{Intra-observation periodicity search}
For completeness we processed the data of the central beam with the \textit{acceleration search} pipeline used to search for pulsars in The High Time Resolution Universe Survey - Northern sky (HTRUN; see \citealt{Barr2013} for survey description), and the \textit{fast folding algorithm} (FFA) based on \citet{Morello2020} implementation. The DM trials ranged from 530 pc cm$^{-3}$ to 590 pc cm$^{-3}$ for the acceleration search, and from 0 pc cm$^{-3}$ to 600 pc cm$^{-3}$ for FFA search. No source was found within the candidates down to a S/N of 8.
\subsection{Green Bank Telescope}
\label{sec:GBT}
We observed FRB~121102 with the GBT's S-Band receiver at a center frequency of 2 GHz and a total bandwidth of 800 MHz, though a notch filter between 2.3 and 2.36 GHz removes interference from satellite radio, slightly reducing the effective bandwidth.  We used the Green Bank Ultimate Pulsar Processing Instrument \citep[GUPPI,][]{GUPPI2009} to coherently dedisperse incoming data at a DM of 560 pc/cm$^{-3}$, recording self and cross-polarization products with 512 frequency channels and a sampling time of 10.24 $\mu$s.\\

We used the PRESTO routine \texttt{rfifind} to flag samples contaminated by interference and applied this mask in subsequent processing \citep{Ransom2011}.  We searched for bursts using the PRESTO routine \texttt{single\_pulse\_search.py}, similar to data from Effelsberg, but with slightly different parameters.  De-dispersed time series were created at trial DMs of 527 to 587 pc/cm$^{-3}$ with steps of 0.2 pc/cm$^{-3}$ and downsampled to a time resolution of 81.92 $\mu$s.  We searched for pulses with a maximum width of 100 ms and inspected all candidates with a S/N $\geq$ 6.  Astrophysical pulses will exhibit a characteristic increase in S/N as the trial DM approaches the true DM of the source.  We also created the dynamic spectrum for each promising candidate to investigate the time and frequency behavior in more detail.  We determined that all high S/N bursts detected in our search were terrestrial interference and did not detect any astrophysical bursts from FRB121102.
\subsection{Arecibo Observatory}
\label{sec:AO}
The simultaneous observations with NuSTAR were obtained through a DDT to the Arecibo Observatory (project code P3219). Note, only half of the scheduled observations were possible due to the telescope shutdown for hurricane Irma. Similarly, simultaneous observations planned along with INTEGRAL were not possible due to the observatory shutdown after hurricane Maria.\\

Data were recorded with the C-lo receiver and PUPPI pulsar backend. PUPPI recorded filterbank files coherently dedispersed to DM$=$557~pc~cm$^{-3}$. The recorded bandwidth at 4.1-4.9 GHz was divided into 512 channels yielding a frequency resolution of 1.56 MHz. The time resolution of the data was 10.24 $\mu$s. The raw data contain full polarization information, which can be used to obtain the rotation measure and polarization profiles in the event of a detection.\\
Before searching, the filterbank data were downsampled in time to 81.92 $\mu$s, the number of channels reduced to 64, and the total intensity (Stokes I) values are extracted. The data were searched with a simple PRESTO based pipeline \citep{Ransom2011}, downsampled in time by a factor of 16 and dedispersed with trial DMs ranging from 507~pc~cm$^{-3}$ and 606~pc~cm$^{-3}$ in steps of 1~pc~cm$^{-3}$. In order to optimize burst detection, the dedispersed time series were convolved with a template bank of boxcar matched filters up to 49~ms. Candidate bursts were identified in the convolved, dedispersed time series by applying a S/N threshold of 6. The resulting diagnostic plots were searched by eye, and no bursts were found. 

\subsection{\nustar}
FRB~121102 was observed by \nustar\ \citep{hcc+13} between 2017 September 5 and 11 in five separate observations with their start times and durations shown in Table~\ref{tab:obs_highenergy}. The data were processed using HEASOFT\footnote{http://heasarc.gsfc.nasa.gov/ftools} and the standard tools \texttt{nupipeline} and \texttt{nuproducts}. We extracted source photons from a $2^\prime$-radius circular region centered on the source position. We used a background region of identical size positioned away from the source. Photon arrival times were corrected to the Solar-System Barycenter using the source position from \citet{Chatterjee2017}.
\subsection{INTEGRAL}
FRB~121102 was observed by the INTEGRAL satellite in late September 2017 in pointing mode and then in 2019 in a Target of Opportunity mode. In both cases, the goal was to search for a possible Hard X-Ray/soft gamma-ray counterpart to the radio emission. The log of the observations with start time and duration is shown in Table~\ref{tab:obs_highenergy}. The INTEGRAL data were processed using the standard INTEGRAL Offline Scientific Analysis (OSA) software\footnote{https://www.isdc.unige.ch/integral/analysis\#Software}. No radio burst was triggered by the Effelsberg telescope during the INTEGRAL exposures, preventing a search for a coincident impulsive event. Searches for correlations before or after the radio burst and with signals from other radio facilities are currently underway (Gouiffes et al, in preparation).


\begin{center}
\begin{table*}
\caption{Technical information of the observation setup for AO, GBT and EFF radiotelescopes listed in Table~\ref{tab:obs}. The frequency refers to the central frequency for the bandwidth of the receiver, T$_{\text{sys}}$ is the temperature system, S/N$_{\text{min}}$ to the minimum detectable signal-to-noise of a single pulse and F$_{\text{min}}$ refers to the fluence threshold for a bursts of 1 ms duration given S/N$_{\text{min}}$.}\label{tab:technical}
\renewcommand{\arraystretch}{1.3}
\begin{tabular}{ C{1.4cm} C{1.4cm} C{1.4cm} C{1.5cm} C{1.0cm} C{1.0cm} C{1.0cm} }
\hline
\hline
Telescope & Backend  & Frequency & Bandwidth & SEFD & S/N$_{\text{min}}$ &  F$_{\text{min}}$\\
& & (GHz) & (MHz)& (Jy) & & (Jy\,ms) \\
\hline
AO & PUPPI & 4.5 & 800 & 3.5 & 10 & 0.03\\
GBT & GUPPI & 2.0 & 740 & 10.0 & 7 & 0.05\\
EFF & PFFTS & 1.36 & 300 & 17.0 & 7 & 0.15\\
\hline 
\end{tabular}
\end{table*}
\end{center}

\section{Burst properties}\label{sec:bursts_prop}
\begin{figure*}
\begin{center}
\includegraphics[trim={0 19.0cm 0.0cm 0 },width=0.9\textwidth]{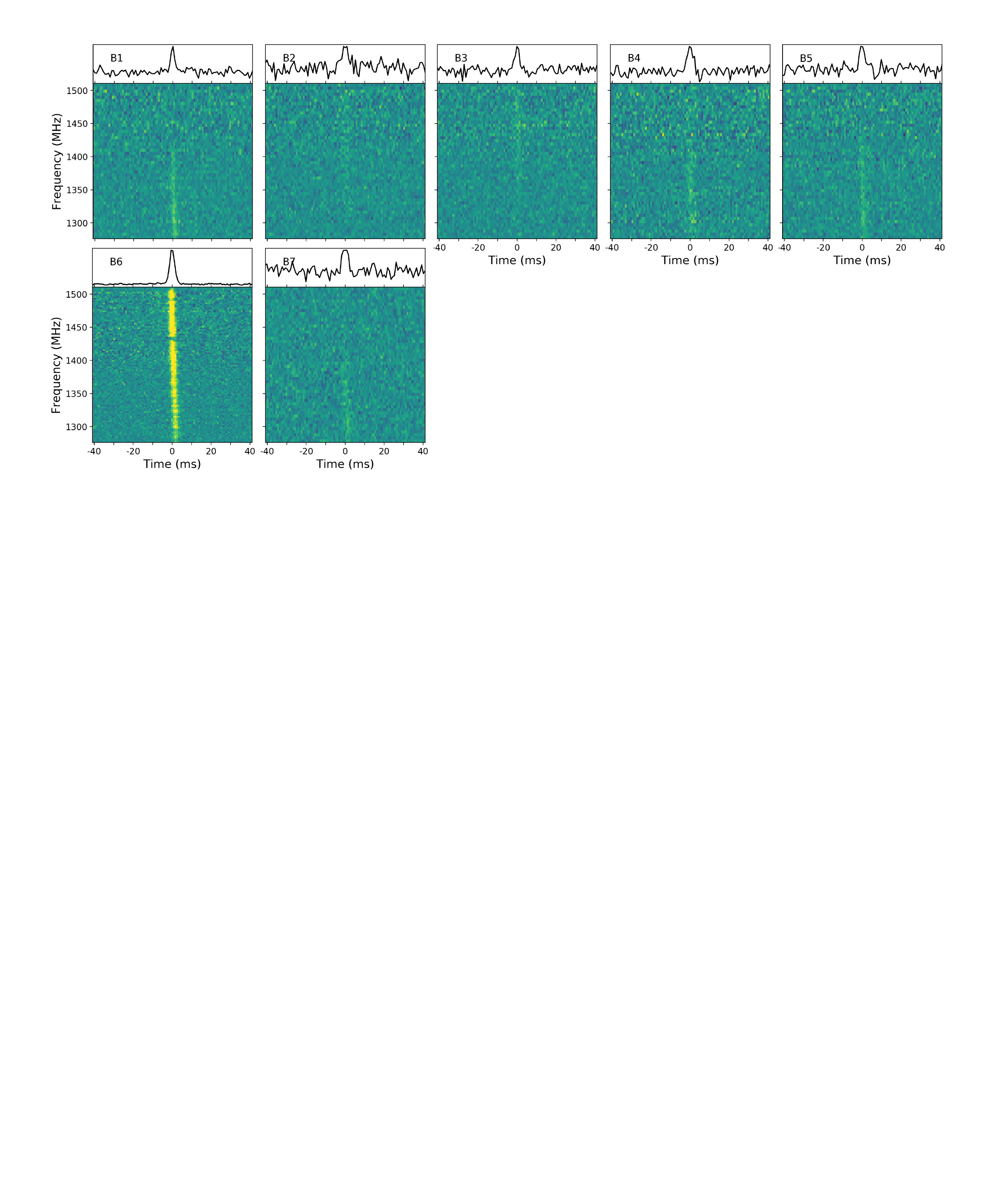}
\caption{Profiles for the bursts detected on September 2017 with Effelsberg at 1.36 GHz. The top plot displays the pulse profiles obtained when integrating in frequency the dynamic spectra in the lower panel. The bursts have been incoherently de-dispersed at 560.5 pc cm$^{-3}$ \citep{Hessels2019}. For visualisation purposes B1--B5 and B7 were frequency scrunched by a factor of 8 and B6 by a factor of 4.}
\label{fig:septemberbursts2017}
\end{center}
\end{figure*}

\begin{figure*}
  
\includegraphics[trim={0 0.5cm 0 0cm},width=0.9\textwidth]{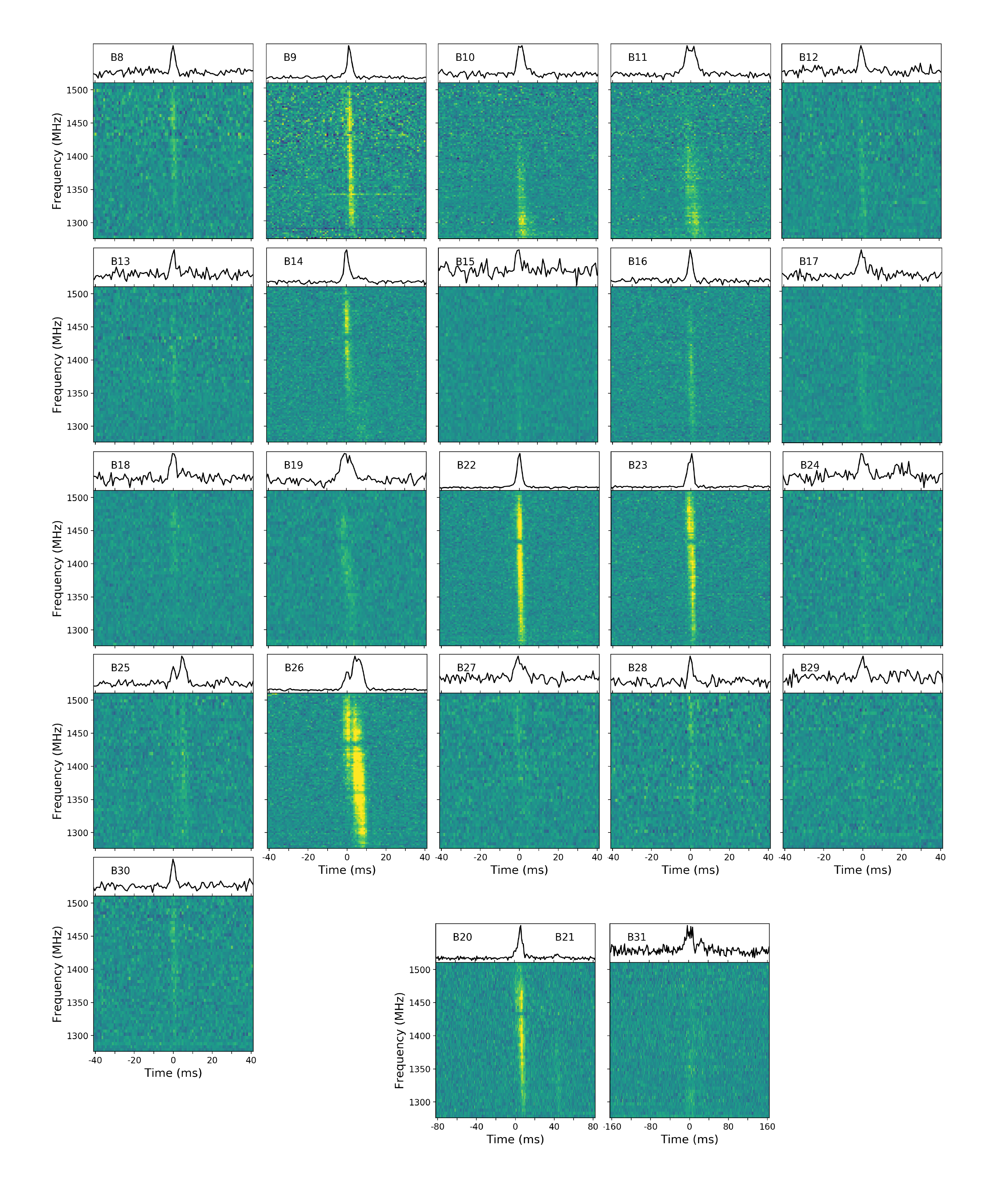} 

\caption{Profiles for the bursts detected on November 2018 with Effelsberg at 1.36 GHz. The top plot displays the pulse profiles obtained when integrating the dynamic spectra in the lower panel. The bursts have been incoherently de-dispersed at 563.5 pc cm$^{-3}$ \citep{Josephy2019}. For visualisation purposes bursts B9--B11, B14, B16, B20--B23 and B26 were frequency scrunched by a factor of 4, B31 by a factor 2, and the rest of the bursts by a factor of 8. }
\label{novemberbursts2018}
\end{figure*}

\begin{figure*}
\begin{center}
\includegraphics[trim={0 24.0cm 0.0cm 0 },width=0.9\textwidth]{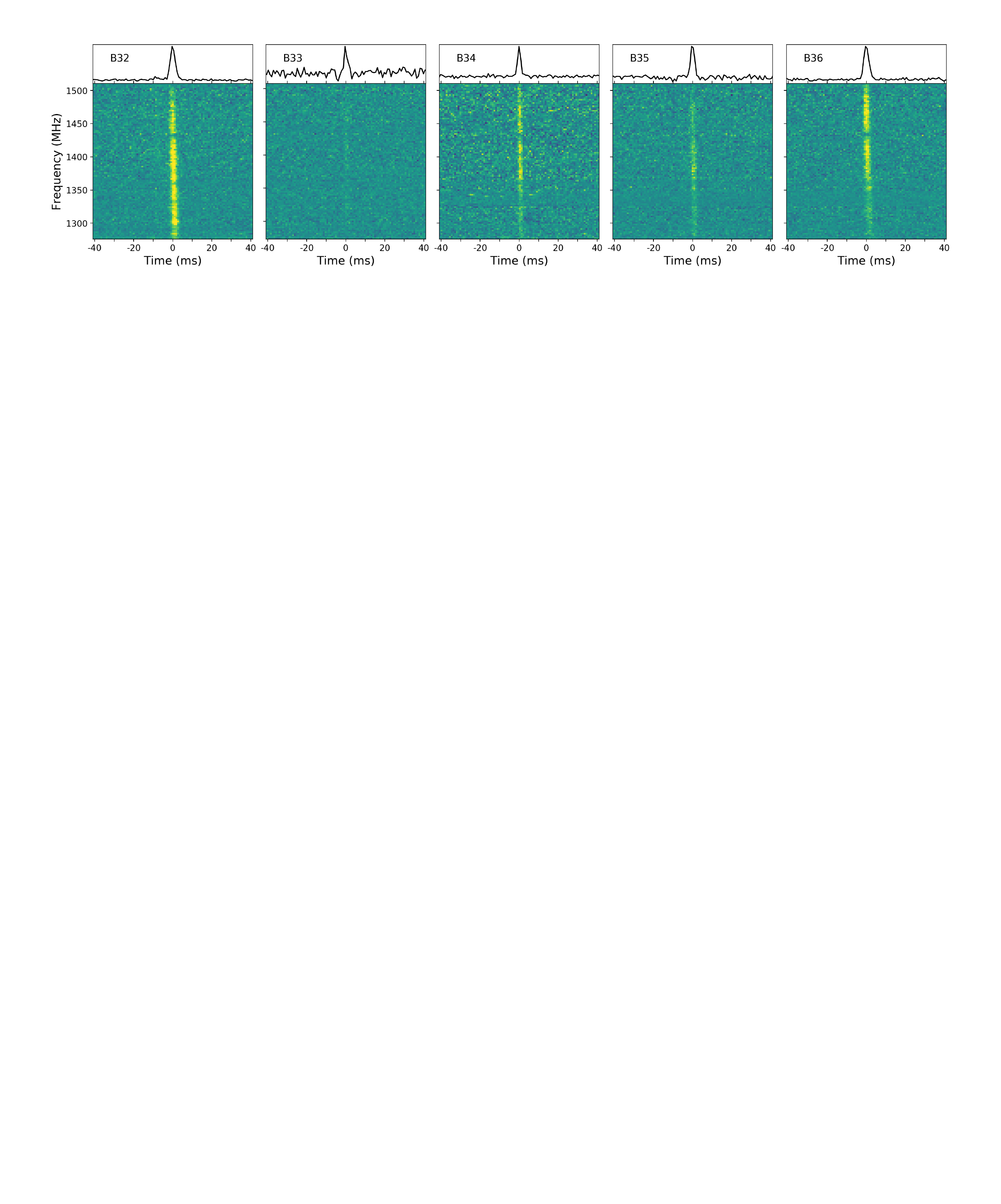} 
\caption{Profiles for the bursts detected on August and September 2019 with Effelsberg.  The top plot displays pulse profile obtained when integrating the dynamic spectra in the lower panel. The bursts have been incoherently de-dispersed at a value of 563.6 pc cm$^{-3}$ \citep{Oostrum2019}. For visualisation purposes all the bursts were frequency scrunched by a factor of 4.}
\label{augustbursts2019}
\end{center}
\end{figure*}

During our follow-up observations, 36 bursts were detected with Effelsberg and none with GBT or AO. The 36 bursts are displayed in Figures~\ref{fig:septemberbursts2017}, \ref{novemberbursts2018} and \ref{augustbursts2019} and their inferred properties in Table~\ref{tab:bursts}. The figures show the dynamic spectra of the bursts over the 235 MHz frequency band, while the top panels show the pulse profiles after integrating the dynamic spectrum in frequency.\\
\\
Our reported bursts come from four epochs (MJD 58002, 58440, 58720, and 58735) and hinting towards three different values for the dispersion measure of the bursts as a result of S/N optimization. However, given the sub-structure present in some of the bursts, such as bursts B11, B20, and B26, we do not report the dispersion value that yields the highest S/N. Furthermore,  we do not attempt to optimize the dispersion value, as our incoherently dedispersed data might not be resolving sub-structure that would otherwise be evident with coherent dedispersion. Instead, we make use of values reported in the literature at epochs near our detections. For the bursts from September 2017 we use the value of 560.5 pc cm$^{-3}$ reported by \citet{Hessels2019}, while for the bursts from November 2018 we use 563.6 pc cm$^{-3}$ \citep{Josephy2019} and for August and September 2019 a value of 563.5 pc cm$^{-3}$ \citep{Oostrum2019}. Such values were deduced from frequency structure optimization by finding the dispersion value that maximizes the forward derivative of the dedispersed time series \citep{Hessels2019}.\\

To calculate the properties of the bursts (Table~\ref{tab:bursts}), such as the S/N, flux density, fluence, and the full-width half-maximum (FWHM), we used DSPSR \citep{DSPSR2011} to extract a snapshot of data from the PFFTS filterbank file and chose a time resolution that matched the intra-channel smearing time. Subsequently, we inferred the properties of each burst as implemented in \citet{Houben2019}. The bursts were fit with a Gaussian-model calculated through least-squares optimization. The height and full width half maximum (FWHM) were obtained afterward from such best fit, and the S/N deduced from the peak and its associated error with the root-mean-square (rms) of the noise fluctuations. We converted these values to flux ($S$) by using the radiometer equation for single pulses: 
\begin{equation}
    S=\frac{(S/N)\cdot \rm{SEFD}}{\sqrt{n_p \cdot \rm{FWHM}\cdot \Delta\nu}}
\end{equation}
\\
for Effelsberg 7-beam's receiver, the system equivalent flux density ($SEFD$) has a mean value of 17 Jy for each of the two polarisations ($n_p$). The bandwidth ($\Delta\nu$) considered for the calculation is 235 MHz, which corresponds to the remaining of the 300 MHz bandwidth for the PFFTS data after cropping the band edges.\\

From Table~\ref{tab:bursts} we see that the brightest bursts of the dataset are B6, B22, B23, B26, with flux densities of 1.56 Jy, 1.4 Jy, 1.0 Jy, and 1.2 Jy respectively, and whose emission extends across the full bandwidth. For bursts such as B10 and B11, their emission comes from the lower part of the frequency band, while B14 is an example of emission coming predominately from the upper part of the frequency band.\\
Six bursts show multi-component profiles: B11, B14, B20, B23, B25, and B26. The number of components ranges from two components, such as B25, to three components for B11 and B26. Particularly interesting bursts are B10 and B14, displaying a weak trailing emission tail, and B20 and B26 with a characteristic downward drifting pattern. B20 and B21 are the closest spaced bursts of our dataset with a TOA difference of $\sim$38 milliseconds. Although the majority of the published bursts have separations above 1 second, it is known that few bursts cluster around TOA separations of  20-40 millisecond \citep{Scholz2017,Hardy2017,Gourdji2019,Caleb2020} and potentially as low as 2.56 millisecond (see \citealt{Gajjar2018}). However, whether this latter case corresponds indeed to two separated events or different components of one burst is ambiguous.\\

B31 is arguably the most interesting burst from the sample. It has a width of 39$\pm$2\,ms, and is, as of now, the broadest burst detected from FRB 121102. To determine its width we have used the FWHM of the Gaussian fit to the two main components of the pulse profile. Given that the known typical durations of the bursts from FRB 121102 are of the order of a couple of milliseconds, we wonder whether the previously mentioned events with separations  20-40 milliseconds \citep{Scholz2017,Hardy2017,Gourdji2019,Caleb2020} correspond to single events in which only the strongest components are detected. This will be discussed further in Section~\ref{sec:discussion}.

\begin{table*}
\caption{Properties of the bursts detected with Effelsberg. The TOAs are barycentred and referred to infinite frequency. All the errors displayed correspond to one standard deviation and arise from the error in the determination of the pulse widths and the rms of noise oscillations in the determination of the S/N}. $S/N_i$ is the signal-to-noise from the discovery as reported by the pipelines and $S/N_f$ is the value obtained after manual cleaning.\label{tab:bursts}
\begin{tabular}{ C{1.2cm} C{2.7cm} C{1 cm} C{1cm} C{3cm} C{2cm} C{2cm} }
\hline
\hline
Burst No & MJD & $S/N_i$ & $S/N_f$  &
Integrated flux density  & Fluence  & FWHM \\
& & & & (Jy) & (Jy ms) & (ms)\\
\hline
B1 & 58002.060649899 & 12.6 &12.1 & 0.19 $\pm$0.02 & 0.47 & 2.4 $\pm$ 0.2 \\
B2 & 58002.060794618 & 6 &7.1 & 0.08 $\pm$ 0.01 & 0.36 & 4.3$\pm$ 0.7\\
B3 & 58002.063269026 & 7.9 &9.2 & 0.13 $\pm$ 0.02 & 0.38 & 2.7 $\pm$ 0.4\\
B4 & 58002.104939709 & 10.4 &10.1 & 0.13 $\pm$ 0.01 & 0.47 & 3.5 $\pm$ 0.4 \\
B5 & 58002.139243447 & 8.8 &9.9 & 0.13 $\pm$ 0.02 & 0.43 & 3.16 $\pm$ 0.4\\
B6 & 58002.166825811 & 93.7 &108.4 & 1.56 $\pm$ 0.2 & 4.64 & 2.97 $\pm$ 0.03\\
B7 & 58002.258539197 & 6.3 &9.2 & 0.12 $\pm$ 0.01 & 0.41 & 3.2 $\pm$ 0.4\\

B8 & 58440.838835667 & 12.1 &13.1 & 0.21 $\pm$ 0.03 & 0.49 & 2.3 $\pm$ 0.2  \\
B9 & 58440.862656975 & 36.1  &38 & 0.58 $\pm$ 0.08 & 1.51 & 2.57 $\pm$ 0.08 \\
B10 & 58440.923733529 & 22 &24.2 & 0.29 $\pm$ 0.04 & 1.20  & 4.0 $\pm$ 0.2 \\
B11 & 58440.936807041 & 29.1 &31.7 & 0.30 $\pm$ 0.04 & 2.04 & 6.7 $\pm$ 0.2 \\
B12 & 58440.989835921 & 11.4 &12.3 & 0.17 $\pm$ 0.02 & 0.53 & 3.0 $\pm$ 0.3 \\
B13 & 58440.998861733  & 7.4  &7.7 & 0.13 $\pm$ 0.02 & 0.27 & 2.0 $\pm$ 0.2 \\
B14 & 58441.007579584 &34.6 &40.8 & 0.58 $\pm$ 0.08 & 1.74 & 2.97 $\pm$ 0.09 \\
B15 & 58441.008221936 & 7 &7.1 & 0.11 $\pm$ 0.01 & 0.26 & 2.4 $\pm$ 0.4 \\
B16 & 58441.012582998 & 18.8 &24.7 & 0.38 $\pm$ 0.05 & 0.96 & 2.5 $\pm$ 0.1\\
B17 & 58441.015540453 & 11.3 &13.8 & 0.11 $\pm$ 0.01 & 1.00 & 8.6 $\pm$ 0.8 \\
B18 & 58441.018047897 &11.1 &11.7 & 0.15 $\pm$ 0.02 & 0.54 & 3.5 $\pm$ 0.4 \\
B19 & 58441.019135880 & 18.1 &22.6 & 0.21 $\pm$ 0.03 & 1.46 & 6.8 $\pm$ 0.4 \\
B20 & 58441.028824049  & 48.4 &56.4 & 0.57 $\pm$ 0.08 & 3.38 & 5.8 $\pm$ 0.1 \\
B21 & 58441.028824494 & 6.4 &8.7 &  0.08 $\pm$ 0.01  & 0.54 & 6.3 $\pm$ 0.9\\
B22 & 58441.029775293 & 84.2 &93.5 & 1.4 $\pm$ 0.2 & 3.84 & 2.73 $\pm$ 0.03 \\
B23 & 58441.059832185  & 66.7 &78.5 & 1.0 $\pm$ 0.1 & 3.57 & 3.35 $\pm$ 0.05 \\
B24 & 58441.059991054  & 7.8 &8.6 & 0.10 $\pm$ 0.01 & 0.43 & 4.1 $\pm$ 0.6 \\
B25 & 58441.061410818 & 14.3 &16.8 & 0.16 $\pm$ 0.02 & 1.07 & 6.6 $\pm$ 0.5 \\
B26 & 58441.064588499 & 105.2 &136.9 & 1.2 $\pm$ 0.1 & 9.38 & 7.6 $\pm$ 0.1 \\
B27 & 58441.067400861 & 8.9 &11.2 & 0.10 $\pm$ 0.01 & 0.70 & 6.4 $\pm$ 0.7 \\
B28 & 58441.106370081 & 9.5 &11.5 & 0.18 $\pm$ 0.02 & 0.42 & 2.2 $\pm$ 0.2 \\
B29 & 58441.109293484 & 7.3 &8.1 & 0.09 $\pm$ 0.01 & 0.41 & 4.2 $\pm$ 0.6 \\
B30 & 58441.112477441 & 10.6 &13.4 & 0.21 $\pm$ 0.03 & 0.51 & 2.4 $\pm$ 0.2 \\
B31 & 58441.124609034 & 14.4 & 20.4 & 0.08 $\pm$ 0.01 & 3.14 & 39 $\pm$ 2 \\
B32 & 58720.206275908 & 77.2 &77.4 & 0.9 $\pm$ 0.1 & 3.13 & 3.20 $\pm$ 0.05 \\
B33 & 58720.230616442 & 8.9 &8.4 & 0.12 $\pm$ 0.01 & 0.34 & 2.7 $\pm$ 0.3\\
B34 & 58735.230804235 & 25.6 &28.3 & 0.49 $\pm$ 0.07 & 1.00 & 2.04 $\pm$ 0.08 \\
B35 & 58735.237756158 & 23.8 &24.2 & 0.38 $\pm$ 0.05& 0.94 & 2.4 $\pm$ 0.1\\
B36 & 58735.239630828 & 46.5 &59.4 & 0.8 $\pm$ 0.1 & 2.61 & 3.13 $\pm$ 0.06 \\
\hline
\end{tabular}
\end{table*}

\subsection{Short-term TOA periodicity search}
Given the high number of bursts detected within one single observation, we searched for an underlying periodicity in the arrival times of the bursts detected in the November 2018 dataset. First, we employed an algorithm commonly used to find the periods of bursts from rotating radio transients\footnote{\texttt{rrat\_period} in PRESTO pulsar search software} \citep{McLaughlin2009}. For a range of trial periods, the number of rotations between consecutive bursts is calculated, and the best period is the greatest common denominator that groups the bursts into the narrowest range of pulse phase. We note that this algorithm does not work well if bursts arrive over a wide range of rotational phases or if there are multiple emission windows. B20 and B21 are only separated by $\sim$38 ms and might be two components of one long burst, so we excluded B21 from our analysis.\\
Using the remaining 23 bursts, this algorithm gave a best-fit period that grouped the bursts within a window of $\sim$0.6 in phase. In order to test the robustness of this finding, we repeated the search 100 times using a sub-sample of 12 randomly chosen bursts. Each trial gave a different period, so we conclude the period determined with all the bursts is not real. Furthermore, we can exclude a period longer than $\sim$0.1 s with a narrow emission window.\\
Lastly, we also carried out a periodogram search over the same 23 TOAs and found no underlying periodicity.
\subsection{\nustar}
\label{sec:Nustar}
The \nustar\ observations overlapped with radio observations, but only one burst was detected while \nustar\ was observing the source. We searched in time near the burst detected by Effelsberg (Burst B6) that occurred during the \nustar\ observation on 2019-09-06. The closest X-ray photon to the time of the Effelsberg burst was 15\,s away. Given this separation, the false alarm probability given the 3--79\,keV \nustar\ count rate of 0.03 count/s is 60\%.\\          
Using an identical method to that employed in \citet{Scholz2017}, we place limits on
the X-ray emission from putative models. That is, we place a count rate limit using the Bayesian method of \citet{kbn91}, and translate that to a fluence limit using the spectral response of \nustar\ and an assumed spectrum. In Table~\ref{tab:nulims} we show the resulting limits.\\       

Compared to the limits placed by \citet{Scholz2017} on X-ray emission at the time of radio bursts from FRB 121102 using {\em Chandra} and {\em XMM}, the limits placed here using \nustar\ are not as constraining for the low absorption case ($10^{22}$\,\nhunits), which corresponds to a typical value for a sightline out of the Milky Way and through the disk of a spiral, Milky-way-like, host galaxy. However, for the highly absorbed case ($10^{24}$\,\nhunits), the \nustar\ limits are about an order of magnitude lower for the hard spectral models (Blackbody and Cutoff PL) and about twice as low for the soft power-law model. These \nustar\ limits therefore further constrain the energetics of X-ray counterparts to radio bursts from FRB 121102, in the case where it is highly absorbed by material close to the source \citep[e.g. supernova ejecta][]{mbm17}.
\begin{table*}
\caption{Burst limits for different X-ray spectral models. The energies are calculated assuming the measured luminosity distance to FRB 121102, 972\,Mpc \citep{Tendulkar2017}. In the table $\Gamma$ is the spectral index. }\label{tab:nulims}
\begin{tabular}{ C{1.5cm} C{1cm} C{1cm} C{2cm} C{2cm} C{2cm} C{2cm} }
\hline
\hline
Model & \nh & kT/$\Gamma$ & Absorbed 3--79 keV & Unabsorbed 3--79 keV & Extrapolated 0.5--10 keV & Extrapolated 10 keV--1 MeV \\  
 & (\nhunits) & (keV/-) & Fluence Limit & Energy Limit & Energy Limit & Energy Limit \\    
 & & & ($10^{-9}$\,\flucgs) & ($10^{47}$\,erg) & ($10^{45}$\,erg) & ($10^{47}$\,erg) \\
\hline
Blackbody & $10^{22}$ & 10  & $2$  & 2   & $8$ & 2 \\
Blackbody & $10^{24}$ & 10  & $3$ & 3 & $12$ & 3 \\
Cutoff PL & $10^{22}$ & 0.5 & $3$  & 3 & $14$ & 50 \\
Cutoff PL & $10^{24}$ & 0.5 & $4$ & 5 & $20$ & 80 \\
Soft PL   & $10^{22}$ & 2   & 0.6  & 0.6 & 60 & 0.9 \\
Soft PL   & $10^{24}$ & 2   & 1.1  & 2 & 180 & 3 \\
\hline
\\[-1.0em]
\multicolumn{7}{l}{\footnotesize{\textbf{Note}: 5-$\sigma$ confidence upper limits.}}\\
\multicolumn{7}{l}{assumed blackbody temperature, kT, for blackbody model and power-law index for power-law models.}
\end{tabular}
\end{table*}
\subsection{Full Effelsberg sample}
In addition to the set of 36 bursts that we report in Table~\ref{tab:bursts}, we incorporate into the following analysis the published datasets from \citet{Hardy2017} and \citet{Houben2019}. Such datasets were acquired with Effelsberg using the identical setup (see Section~\ref{sec:Eff}). In total, our sample contains 57 bursts detected in 165 hours of observations in epochs between MJD 57635 to 59006.  
\subsubsection{Long-term periodicity}\label{subsec:periodicity}
We create a time series from the dates of the EFF observations and label a session with 1 when at least one event was detected and with 0 when no bursts were detected.  Because the observations were not done with a regular cadence, the time series is unevenly sampled. We search for a periodic signal through a periodogram analysis.\\

The majority of the observations were not triggered based on known source activity, but instead, scheduled based on the availability of the higher energy telescopes. The exception is the observations in early September 2017 coordinated with {\emph NuSTAR}, which were scheduled based on the GBT detections at C-band presented in \citet{Gajjar2018}.
Furthermore, we consider a periodogram search to be a valid approach, as the full sample of detections and non-detections is included. Note, our dataset, which totals 34 epochs, is composed of roughly 70\%  non-detections. This is often not the case, as published data is biased towards detections.\\

\begin{figure*}
\begin{center}
\includegraphics[width=0.95\textwidth]{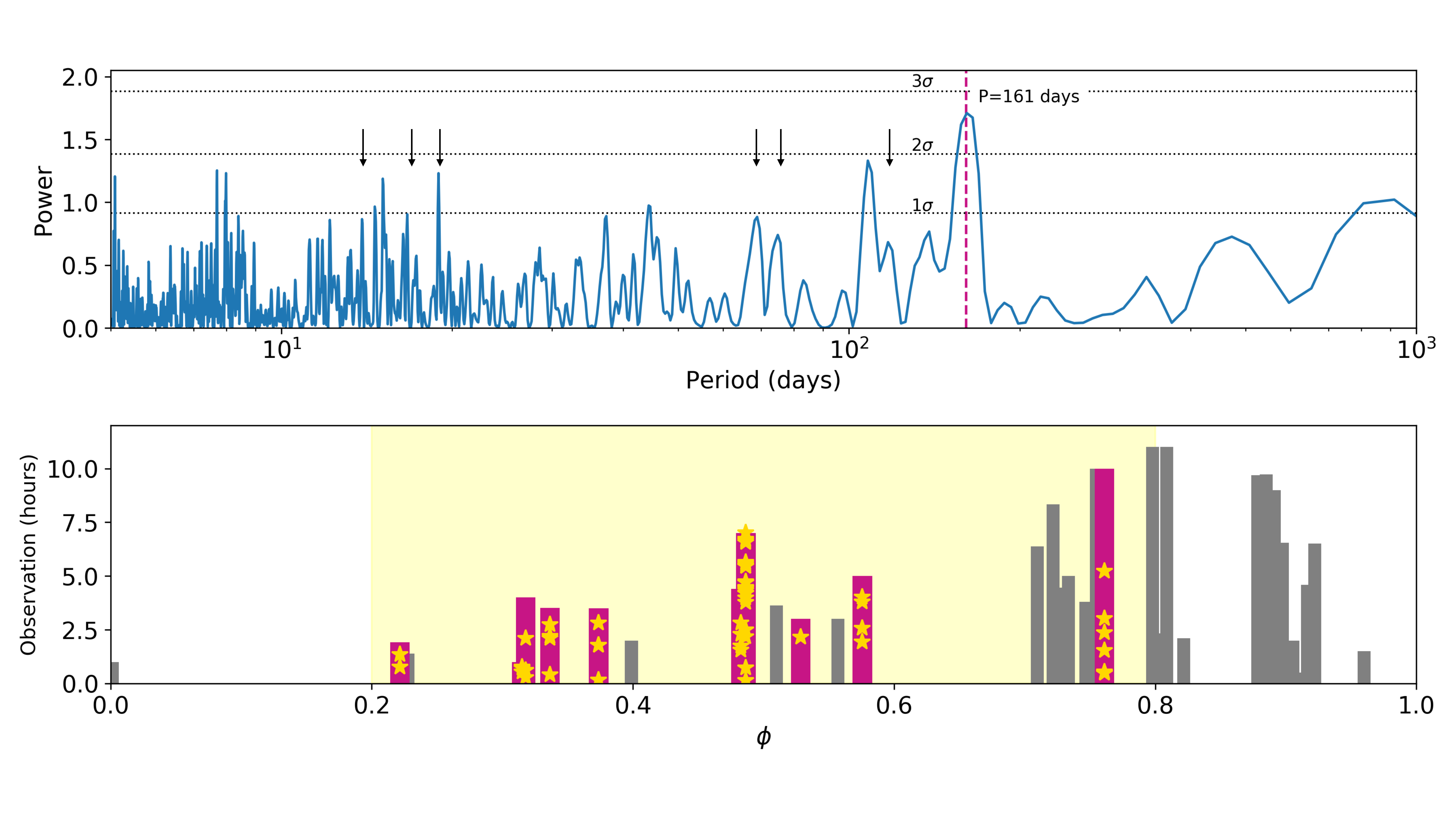} 
\caption{Periodicity analysis for FRB121102. Top: Lomb-Scargle periodogram for the Effelsberg dataset at 1.36 GHz composed of 34 epochs from September 2016 to June 2020. The vertical dashed line shows the best period prediction and the arrows show peaks coming from the window transform. The horizontal dotted lines show the 1-$\sigma$, 2-$\sigma$, and 3-$\sigma$ significance levels deduced from 10,000 bootstrap resamplings. Bottom: phases of the observations based on a 161 days periodicity displayed against the length of its observation. In Magenta are highlighted the epochs with detections for which the yellow-\textit{stars} indicate the time within a given observation where the bursts occurred. The bars in grey are the observations for which no bursts were detected, and the yellow-shaded region is the estimated active phase from the L-band dataset and referred to MJD 57075 as the epoch with phase $\phi=0$.}
\label{fig:periodicity}
\end{center}
\end{figure*}

Following the formulation presented in \citet{LS2018}, we proceed with the Lomb-Scargle periodogram, which is displayed on the top of Figure~\ref{fig:periodicity}. First, we subtract the mean value from the time series. This step is important as the Lomb-Scargle model assumes that the data is centered around the mean value of the signal. As seen in Figure~\ref{fig:periodicity} the periodogram peaks at a period of 161$\pm$5 days, in agreement with the postulate of \citet{PeriodicityFRB121102}. The 1-$\sigma$ uncertainty is not estimated from the width of the peak, as this is not optimal for time series with long baselines and few data points. Instead, we determine the uncertainty of the peak, $\sigma_{ls}$, through \citep{LS2018}:

\begin{equation}
\sigma_{ls} = \frac{\textrm{FWHM}}{2}\sqrt{\frac{2}{\textrm{N}\cdot\textrm{S/N}^2}}
    \label{eq:sigma}
\end{equation}
where N is the number of points in the dataset and FWHM is the full-width at half-maximum of the Gaussian fit to the peak.\\ 

Given the presence of several peaks with significant power in the Lomb-Scargle periodogram, we investigate if any are introduced from the observing function. This is plausible as the Lomb-Scargle periodogram of the data in Figure~\ref{fig:periodicity} is the result of the convolution of the true signal from FRB 121102, and a set of top-hat functions with different durations, which describe the observations (window function). We compute the Lomb-Scargle periodogram of the window function by keeping the epochs unchanged but setting all values to one (for detections and non-detections). For the window transform the data is not centered. We identify several peaks in the periodogram of the window function that are also present in the periodogram of the data. Some peaks among the top 15 periods are marked in Figure~\ref{fig:periodicity} with the black arrows and correspond to (from right to left) 119, 75, 70, 19, 16, and 14 days roughly. More importantly, the analysis of the window function down to the top 20 periods did not show any peak at 161 days, supporting the conclusion that the 161 day period is in the data.\\

We further test the hypothesis of an underlying periodicity through the use of the bivariate \textit{l$_1$}-periodogram, also referred to as the \textit{robust-}periodogram \citep{RP_LI2010}. This type of periodogram is derived from the maximum likelihood of multiple frequency estimation, and it uses the least-absolute deviations regression model - instead of the least-squares minimization as the Lomb-Scargle periodogram -  which is a robust regression against heavy-tailed noise and outliers. The robust periodogram predicts a 161 days period as well, perfectly in agreement with the Lomb-Scargle prediction.\\
\\
To calculate the significance of the peak, we estimate the false alarm probability using a bootstrap method with 10,000 trials. We keep the epochs unchanged and for each trial draw randomly the outcome of an observation (detection or non-detection). We record the maximum power of each generated Lomb-Scargle periodogram and calculate the probability that a given power exceeds a threshold through percentile rank. The dotted lines in the Lomb-Scargle periodogram in Figure~\ref{fig:periodicity} show the 1-$\sigma$, 2-$\sigma$, and 3-$\sigma$ significance levels for the highest peak, determined by the 10,000 bootstrap resamplings. We determine a significance of roughly 2.7-$\sigma$ for the 161 days peak. This approach answers the question of how likely it is that any period will have, by chance, a power above a given value. However, this is a conservative approach, and in the case of non-Gaussian noise, it underestimates the significance levels.\\
\\
We ask now, specifically, how likely it is that a period of 161 days, by chance, will have a signal power above 1-$\sigma$, 2-$\sigma$ and 3-$\sigma$ significance levels. This is equivalent to a false-positive rate of 161 days period among 10,000 bootstrap trails. We run the simulation and keep the powers encountered at 161 days. Through this approach, the peak is found to have a significance above 4-$\sigma$ level. We do not approach more sophisticated methods to estimate the false alarm rate as it is outside the scope of the paper, but we clearly show that the periodicity reported by \citet{PeriodicityFRB121102} is also seen in our dataset.\\

In the lower panel of Figure~\ref{fig:periodicity}, we see the outcome of assigning a phase to each epoch by folding at a period of 161 days. The y-axis shows the length of each observation. We take MJD 57057 as reference for phase $\phi=0$. From the outermost observations with detections, we infer an active window of 54\%. However, we notice that while the end of the active window is densely sampled (by chance), the start of the active phase is not. Motivated by this, and in order to test how representative the Effelsberg dataset is, we add the published follow-up observations on FRB 121102 at L-band (1-2 GHz). The frequency constraint is driven by fact that simultaneous observations at radio frequencies greater than $\sim$GHz have mostly not led to simultaneous detections, suggesting that a given activity extends only over a couple of hundreds of Megahertz \citep{Law2017}. We extend the dataset by including the detections and non-detections reported by \citet{Spitler2014,Spitler2016}, \citet{Scholz2016,Scholz2017}, \citet{Gourdji2019} and \citet{Oostrum2019}. We refer hereon to this dataset as the L-band dataset.\\

The L-band dataset is composed of a total of 179 epochs from which 43 are detections and 136 non-detections. The Lomb-Scargle periodogram for this dataset infers a period of 158$\pm$3 days, which is consistent with the period deduced independently from the EFF dataset and with \citet{PeriodicityFRB121102}. To define a more constraining active phase we proceed with the prediction of 161 days and fold the L-band epochs. By considering the left and right-most observations with detection in the L-band dataset we define an active phase of roughly 60\%, which is shown in Figure~\ref{fig:periodicity} with the yellow-shadowed region. Based on the inferred periodicity of 161 days and the active phase of 60\%, we construct the active windows. We use as reference for $\phi\,=\,0$ the MJD 57075, and find that the epochs with detections fall into 5 activity windows: 57590-57687, 57751-57848, 57912-58009, 58395-58492 and 58717-58814.\\

We note that the width of the active phase is dependent on the selection of the period and on observations at the start and end of the on-phase. For periods between 156 days and 161 the active window ranges from 56\% to 62\%. It is worth noting that \citet{PeriodicityFRB121102} defined the period from dispersion minimization, i.e. choosing the period gives the narrowest possible active window. Naturally, more observations with detections outside of the limits defined here will broaden the active window.\\
\subsubsection{Repetition pattern on active phase}\label{subsec:weibull}

In this section, we investigate the waiting time statistics between consecutive bursts on shorter time scales. Time independent Poissonian statistics as well as Weibull distribution with shape parameter $k<1$ have been previously assumed. While $k<1$ means that clustering is present in the data (the lower the $k$ the higher the degree of clustering), when $k=1$ the Poissonian case is recovered, and for $k>>1$ we are in the presence of a constant separation, implying periodicity.\\

\citet{Oppermann2018} used a Weibull distribution with shape parameter $k$ smaller than one using $\sim$80 hours of FRB 121102 follow-up data. In the analysis of \citet{Oppermann2018}, the sample contained observations taken with Arecibo, Effelsberg, GBT, VLA, and Lovell at different observing frequencies ranging from 0.8 - 4.8 GHz. The 17 bursts contained in the data led to the conclusion that a Weibull distribution with a shape parameter of  k=0.34 and a mean event rate of r=5.7 $\mathrm{day}^{-1}$, is a much better descriptor for the time interval between consecutive events than Poissonian statistics. Recently, \citet{Oostrum2019} came to a similar conclusion using WSRT/Apertif data.\\

We would like to test whether this strong clustering observed was a consequence of the unknown periodicity of FRB 121102. However, in a different approach to the one carried in \citet{Oppermann2018} work, we do not combine bursts from different telescopes and observing frequencies, as their difference in sensitivity leads to different event rates, and this might bias the observed clustering. As mentioned in Section~\ref{subsec:periodicity} our observations fall into 5 activity windows: MJD 57590-57687, 57751-57848, 57912-58009, 58395-58492 and 58717-58814. We group all the observations with and without detections falling into such windows to study the waiting time between consecutive bursts.\\

\begin{figure*}
\begin{center} 
\includegraphics[width=\textwidth]{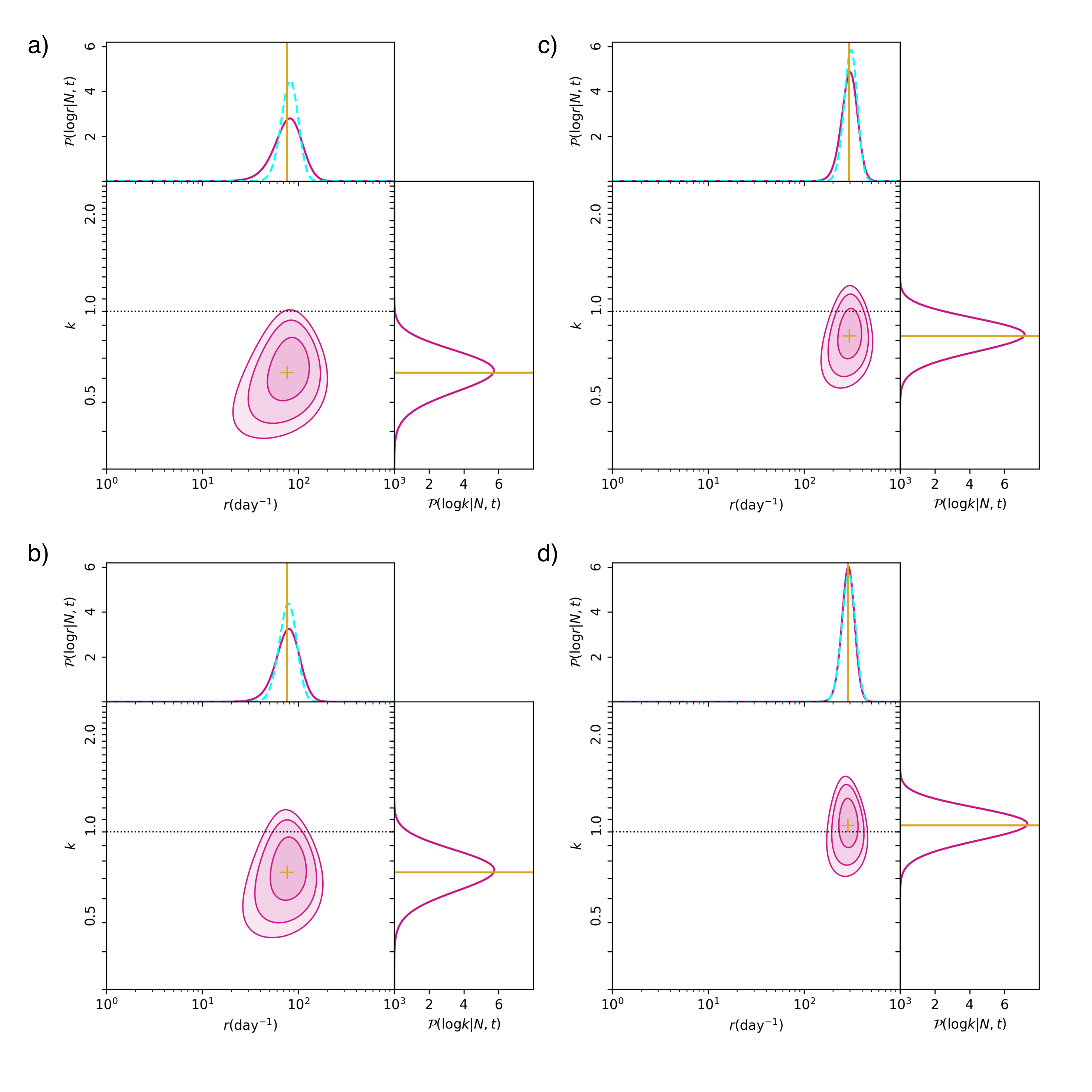} 
\caption{2-dimensional posterior probability distribution for the shape parameter \textit{k} and the event rate \textit{r} for a Weibull distribution (magenta). The cyan-dashed curve represents the expectation for the event rate from a classic Poissonian distribution and the yellow lines indicate the mean values of the posterior distribution for \textit{k} and \textit{r}. The contours in the parameter space represent 65\%, 95\% and 99\% confidence intervals, and the horizontal dotted-lines mark $k=1$ for reference. Top: fit to all the events in November 2018 sample (panel a) and for \citet{Gourdji2019} in panel (c). Bottom: fit to November 2018 set (panel b) and \citet{Gourdji2019} (panel d) when excluding events with $\delta t<1$ s respectively.}
\label{fig:2Dposterior}
\end{center}
\end{figure*}
We use the two-dimensional posterior probability for $k$ and $r$, and the one-dimensional marginal posterior for $k$ and $r$ formalism implemented by \citet{Oppermann2018}\footnote{https://github.com/nielsopp/frb\_repetition}. Our first approach is to treat each active window independently. However, this implicitly assumes that the event rate is constant across the full active phase. Clearly, this must not be the case, for instance, if the on-window has a Gaussian profile, which could lead to higher event rates at the center of the window. Nonetheless,  except for the epoch of November 2018, we note that there are not sufficient bursts per active window to reduce the parameter space of the posterior probability to well-constrained values. This means that it cannot be differentiated between a Poisson and Weibull distribution. Because the November 2018 observation consists of a single 7-hour long session, we focus now on the statistics within a single, long observation.

The result for the November 2018 epoch is shown in Figure~\ref{fig:2Dposterior} (left) and the values for $k$ and $r$ presented in Table~\ref{tab:posterior_val}. The posterior distribution for all the bursts of November 2018 in panel a) Figure~\ref{fig:2Dposterior} shows that for a Weibull fit the shape parameter is $k=0.62^{+0.1}_{-0.09}$ and the event rate is $r=74^{+31}_{-22}$\, day$^{-1}$ (magenta curves in Figure~\ref{fig:2Dposterior}). From a Poissonian distribution the average rate is  $r_{\rm{p}}=82\pm27$\, day$^{-1}$ (cyan curves in Figure~\ref{fig:2Dposterior}). Both values report 1-$\sigma$ intervals and consider a fluence threshold of 0.15 Jy\,ms for bursts of 1\,ms duration which is imposed by Effelsberg's sensitivity to bursts with S/N > 7.\\\

\begin{figure}
\begin{center} 
\includegraphics[width=0.5\textwidth]{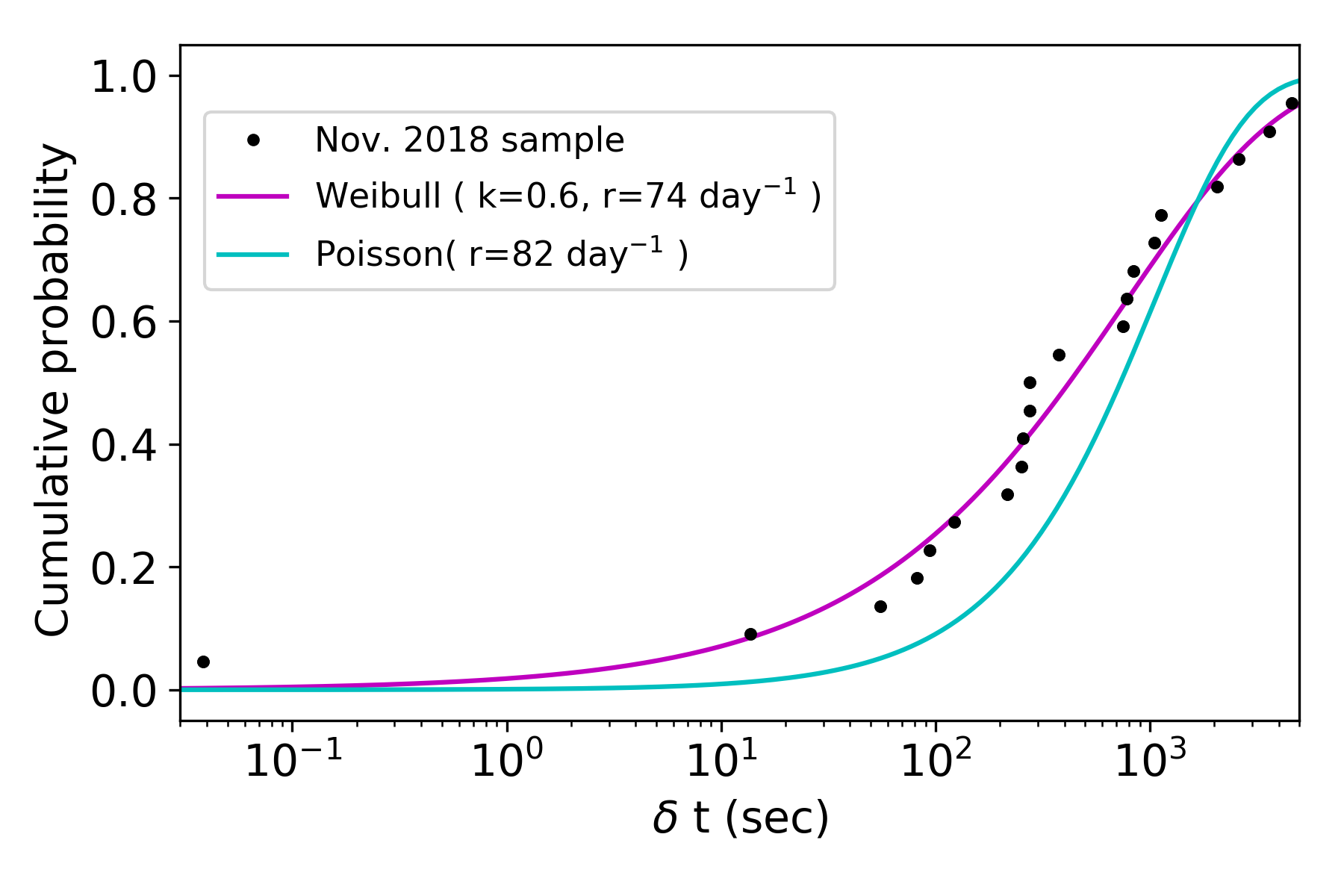} 
\caption{Empirical cumulative density function of the waiting time between consecutive bursts ($\delta$t) of FRB 121102 for the November 2018 dataset (bursts B8 to B31 in Table~\ref{tab:bursts}). The magenta and cyan color represent the best-fit from the Weibull and Poisson cumulative density functions, respectively.}
\label{fig:pdf}
\end{center}
\end{figure}

To perform an additional test on how well Poisson and Weibull distributions fit this set of bursts, we plot the empirical cumulative density function (ECDF) of the time interval between consecutive bursts ($\delta t$) in Figure~\ref{fig:pdf}. The cumulative density functions from Weibull ($P_{\rm{w}}$) and Poissonian ($P_{\rm{p}}$) statistics are described by \citep{Oppermann2018}:\\
\begin{equation}\label{eq:Weibull_eq}
P_{\rm{w}}\,\left(\delta t,k,r\right) = 1-e^{-(\delta t\,r\Gamma(1+1/k))^k)}
\end{equation}
\begin{equation}\label{eq:Poisson_eq}
P_{\rm{p}}\,\left(\delta t,r_p\right) = 1-e^{-\delta t\,r_p}
\end{equation}
in the equations above $r_p$ and $r$ represents the event rates for Poissonian and Weibull case, respectively, $k$ is the shape parameter and $\Gamma$ is the incomplete gamma function. $k$ and $r$ are taken from the mean value of the two-dimensional posterior probability function.\\

Figure~\ref{fig:pdf} shows the ECDF of the November 2018 dataset alongside the fit from the Weibull and Poisson models. Qualitatively, Weibull's CDF better describes the distribution. To quantify the fit we compute the Kolmogorov-Smirnov statistic (KS test) of the November 2018 sample versus its Weibull distribution fit and versus its Poisson distribution fit. We obtain a p-value of 0.84 and 0.12 for the Weibull and Poisson fit respectively, meaning that the null hypothesis - of the November sample drawn from a Weibull or Poisson distribution - cannot be rejected. Nonetheless, when considering as well the KS statistic we find that the absolute max distance between the ECDF  of the sample and the CDF fits (equations \ref{eq:Weibull_eq} and \ref{eq:Poisson_eq}) are 0.13 and 0.27 for the Weibull and Poisson fit respectively. Therefore, we conclude that for the full November 2018 sample the best fit is a Weibull distribution with $k=0.62^{+0.1}_{-0.09}$ and the event rate of $r=74^{+31}_{-22}$\, day$^{-1}$.\\
It is worth noting that the tail to the left of the ECDF in Figure~\ref{fig:pdf} is fit by neither of the distributions. This single event corresponds to the separation of ${\sim}38$ ms between bursts B20 and B21, as previously discussed. From a Poisson distribution, the probability of having a waiting time of 38 milliseconds or shorter is 0.003\%. We recall the previous discussion on B31 and its 39$\pm$2\,ms duration, and the cluster behavior for bursts with waiting times shorter than 1 second observed by \citet{Li2019} and \citet{Gourdji2019}. Therefore, we exclude events with $\delta\,t<$1\,s to explore whether this single wait time has a strong effect on the determination of the shape parameter for the Weibull distribution. We obtain $k' = 0.73^{+0.12}_{-0.10}$ and show in Figure~\ref{fig:2Dposterior} panel b the posterior distributions after excluding $\delta\,t<$1\,s. It is observed that removing a single wait time narrows the posterior distribution for the event rate and moves it closer to a Poissonian distribution.\\
\\
To expand on this strong dependence of the shape parameter on a few clustered events, we additionally model independently the 41 bursts detected with Arecibo by \citet{Gourdji2019}. We find $k = 0.82^{+0.1}_{-0.09}$ for the whole 41 bursts (panel c Figure~\ref{fig:2Dposterior}) and $k' = 1.0^{+0.2}_{-0.1}$ for the resultant 39 bursts when excluding events with $\delta t<1$\,s (panel d Figure~\ref{fig:2Dposterior}). For this dataset, the exclusion of 2 bursts results in a change to the Poissonian case. Interestingly the sub-set of bursts from our Effelsberg sample from \citet{Houben2019} includes two epochs separated by a couple of days before and after the \citet{Gourdji2019} detections. The posterior probability for \citet{Houben2019} subset gives a $k = 1.0^{+0.4}_{-0.2}$, which agrees with the $k$ value that we obtained for  \citet{Gourdji2019}.\\
\\
We discuss the meaning of the exclusion of events with  $\delta\,t<1$\,s in Section~\ref{sec:discussion}.

\subsubsection{Energy distribution}\label{subsec: energy_distr}
\begin{figure*}
\begin{center}
\includegraphics[width=\textwidth]{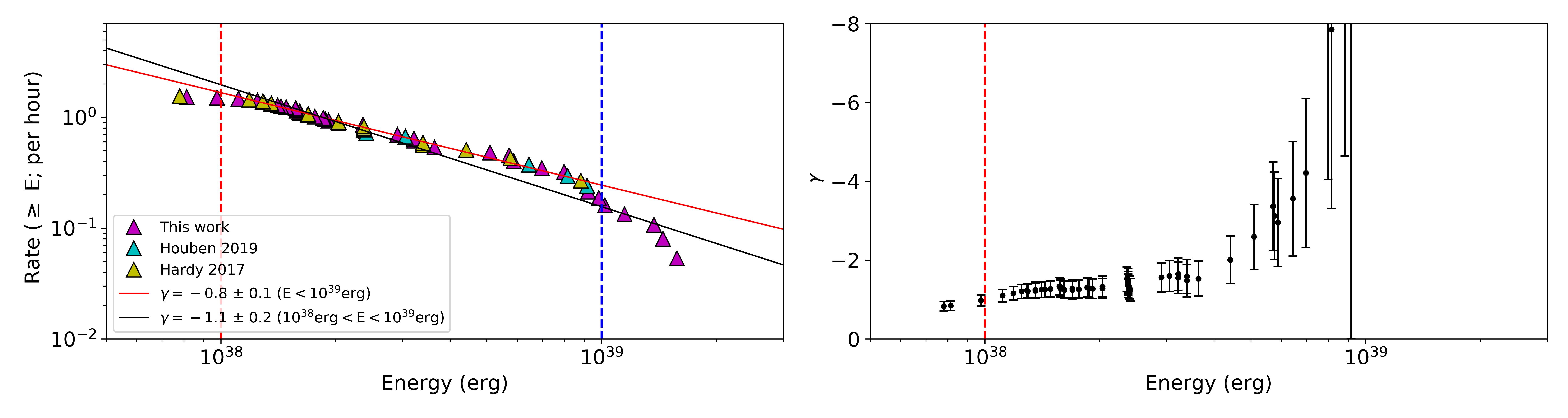} 
\caption{Left: cumulative energy distribution of the bursts from FRB 121102 detected by Effelsberg at 1.36 GHz, shown in magenta for the dataset presented in Table~\ref{tab:bursts}, in yellow for the detections from \citet{Hardy2017} and in cyan for \citet{Houben2019}. The isotropic energy is calculated as described in equation (\ref{eq:energy}) and the fit corresponds to a power-law of the form $N \propto E^\gamma$ estimated through maximum-likelihood. The red-dashed line shows the completeness limit for Effelsberg at $\sim$E$=10^{38}$ erg and the blue-dashed line marks the bursts above the saturation limit roughly at E$=10^{39}$ erg. Right: convergence of the power-law slope $\gamma$ as determined from the Maximum-likelihood estimation. The red-dashed vertical line represents the truncated value by excluding bursts below the completeness limit.}
\label{fig:energy_dist}
\end{center}
\end{figure*}
In this section, we study the isotropic energy distribution of the 57 bursts that compose the Effelsberg sample as it can provide insights on the mechanisms responsible for its emission generation. Some sources with accretion disks, such as X-ray binaries and AGNs show a log-normal relation in their flux distribution \citep{kunjaya2011}, other sources, such as high-energy bursts of magnetars \citep{Gogus2000} and the X-ray flares from Sgr A$^{\star}$ \citep{Li2015}, show a power-law in the form of $N \propto E^\gamma$, where $N$ is the rate of events above a given energy value $E$. Interestingly, sources like pulsars show a bimodal relation: their regular emission is well modeled with a log-normal distribution \citep{Burke-Spolaor2012}, while giant pulses are modeled by a power-law distribution \citep{Ramesh2010}. Examples of sources exhibiting a power-law behaviour for their energy distribution are the Crab pulsar, whose giant pulses distribute with an index $\sim\gamma=-2$  \citep{Popov2007,Ramesh2010} and high energy bursts from magnetars with index $\gamma=$-0.6 -- -0.7 \citep{Gogus1999,Gogus2000}.\\

Given its repeating nature, FRB 121102 allows us to study how its bursts distribute over a given energy range. This energy range is so far limited to radio frequencies. We compute the isotropic energy $E$ for a given burst as:\\
\setlength{\abovedisplayskip}{15pt} \setlength{\abovedisplayshortskip}{0pt}
\begin{equation}
E=\dfrac{1}{1+z}\,F \textrm{(Jy\,s) }\times\Delta\nu \textrm{(Hz)}\times10^{-23}\mathrm{ergs^{-1}\,cm^{-2}Hz^{-1}} \times 4\pi\,L^2
    \label{eq:energy}
\end{equation}
where $z$ is the redshift, $F$ is the fluence of the burst, $\Delta\nu$ is the bandwidth and $L$ is the luminosity distance, whose value we take as 972 Mpc   \citep{Tendulkar2017}. Figure~\ref{fig:energy_dist} shows the cumulative energy distribution of all the bursts, which is calculated from the cumulative distribution of the mean burst rate. The plateau observed for energies below $10^{38}$ erg, is due to the reach of the completeness limit, i.e threshold where we are not sensitive enough and start missing events.\\

We fit a slope to the data using the Maximum-likelihood method for a power-law fit as described in \citet{James2019} and obtain a slope of $\gamma=-1.1\pm 0.1$ when we exclude the bursts below the completeness level and above of the saturation. The saturation is caused by the down-conversion of the data from 32-bits to 8-bits.\\

We tested the saturation limit by analyzing single pulses from B0329+54, one of the brightest pulsars in the northern sky. We compared the SNR of the bursts in the raw 32-bit data and in the converted 8-bit data. We concluded that low-to-mid SNR bursts showed a nearly one-to-one relation, but as the S/N exceeds $\sim$80 the relation starts breaking leading to a drop in the SNR. We have determined the saturation limit to be affecting bursts above energies of $10^{39} \mathrm{ergs}$ based on the expected energy of a burst with an SNR of 80 and FWHM of 5\,ms (average duration from our sample). 

On the lower end of the energy range, the completeness threshold has been defined based on the convergence of $\gamma$ shown at the right of Figure~\ref{fig:energy_dist}. Each data point corresponds to the determination of $\gamma$ from the inclusion of a successive burst. As the bursts are sorted from higher to lower energy the convergence of $\gamma$ goes from the right to the left. We see in Figure~\ref{fig:energy_dist}, that if the last three bursts with energies below $10^{38} \mathrm{ergs^{-1}}$ are not considered, the slope of the power-law fit is truncated to $\gamma=-1.1\pm 0.1$.

\begin{center}
\begin{table*}
\caption{Posterior values for the event rate ($r$) and shape parameter ($k$) from a Weibull Statistics and the event rate $r_p$ from a Poissonian distribution. Prime values indicate the outcome of excluding bursts with waiting times shorter than 1 second ($\delta t<1$\,s). The event rates consider a fluence threshold 0.15 Jy\,ms for bursts with a 1 ms duration and S/N above 7, and the confidence intervals shown assume 1-$\sigma$ uncertainties.}\label{tab:posterior_val}
\renewcommand{\arraystretch}{1.3}
\begin{tabular}{ C{3cm} C{1cm} C{1cm} C{1.2cm} C{1cm} C{1cm} C{1.2cm} }
\hline
\hline
Dataset & $r_p$ & $r$ & $k$ & $r_p'$ & $r'$ & $k'$\\
& (day$^{-1}$) & (day$^{-1}$) & & (day$^{-1}$) & (day$^{-1}$) &\\
\hline
All  & $8\pm3$ & $7^{+3}_{-2}$ & $0.40^{+0.04}_{-0.03}$ & $8\pm3$ & $7^{+3}_{-2}$ & $0.43^{+0.04}_{-0.03}$\\
On-$\phi$ & $18\pm4$ & $18^{+5}_{-4}$ & $0.50^{+0.05}_{-0.03}$ & $17\pm4$ & $17^{+5}_{-3}$ & $0.57^{+0.06}_{-0.04}$\\
November 2018  & $82\pm9$ & $74^{+31}_{-22}$ & $0.62^{+0.10}_{-0.09}$ & 79$\pm9$ & $76^{+26}_{-19}$ & $0.73^{+0.12}_{-0.10}$\\
\citet{Gourdji2019} & 307$\pm17$ & $294^{+57}_{-52}$ & $0.82^{+0.12}_{-0.09}$ & 292$\pm17$ & $286^{+44}_{-44}$& $1.0^{+0.2}_{-0.1}$\\ 
\citet{Houben2019}  & 20$\pm4$ & $18^{+8}_{-5}$ & $1.0^{+0.4}_{-0.2}$ & - & - & - \\ 
\hline
\end{tabular}
\end{table*}
\end{center}
\section{Discussion}\label{sec:discussion}
It is long known that detections of FRB 121102 are clustered in time. Much of this clustering likely reflected a not-yet-defined periodicity. Here we investigated whether there is still clustering within a single observation or active phase.\\

We defined active windows for FRB 121102 based on the 161 days period deduced from the 34 total epochs in the Effelsberg dataset but constrain the width of the active phase to be roughly 60\% based on the 179 observations from the L-band band dataset. This mixed approach is motivated by the poor coverage of the phases prior to the start of the active phase as seen in Figure~\ref{fig:periodicity}. This phenomenon is purely by chance as the observations were scheduled mostly based on telescope availability, which led to unevenly sampled epochs and different observation lengths. We use the MJD 57075 as a reference epoch and estimate the next active phase to be from MJD 59039 to 59136 followed by a period of inactivity until the next cycle from MJD 59200 to 59297. Our predicted active phase is shifted with respect to the prediction from \citet{PeriodicityFRB121102} by 37 days later, due to the difference in periodicity of 157 and 161 days predicted by \citet{PeriodicityFRB121102} and our work, as well as the estimated on-phase of 54\% and roughly 60\% respectively, and ultimately due to the difference in the reference epochs used of MJD 58200 in \citet{PeriodicityFRB121102} and MJD 57075 used by us. This is not surprising as the entire phase is not fully sampled by the existing observations. Besides observations during the predicted window, where chances to detect bursts are higher, we also recommend unbiased observations through the entire phase to better constrain the activity window.\\

Regarding the waiting time between consecutive bursts, we first consider all the Effelsberg data - composed of 165 hours of observations and 57 bursts - and ask how clustered the events are. From the two-dimensional posterior probability for \textit{k} and \textit{r} and the one-dimensional marginal posterior for \textit{k} we obtain an event rate of  $7^{+3}_{-1}\rm{day}^{-1}$ and \textit{k}=$0.40^{0.04}_{0.03}$ for a Weibull distribution and a rate of  $8\pm3$ day$^{-1}$ from Poisson. On the other hand, if we acknowledge the presence of a periodic active phase and restrict the observations to the 5 active windows in which the Effelsberg data falls: MJD 57590-57687,  57751-57848,  57912-58009,  58395-58492, and 58717-58814 we determine $18^{+5}_{-4}\rm{day}^{-1}$ and \textit{k}=$0.50^{0.05}_{0.03}$ for a Weibull distribution and a rate of  $18\pm4$ day$^{-1}$ from Poisson. For both datasets, we also calculate the values after excluding wait times with $\delta t<1$\,s. All values are listed in Table~\ref{tab:posterior_val}. The rates consider events above a fluence threshold of 0.15 Jy\,ms for bursts of 1\,ms duration which is imposed by Effelsberg's sensitivity to bursts with S/N > 7.\\

However, the previous results for \textit{k} and \textit{r} when restricting the observations to the active window implicitly assume that the event rate is constant across all the active windows and across the full active phase, which may not be the case. If we restrict the analysis to each active window independently, we notice that November 2018 sample is the only one providing sufficient bursts in a single active window to reduce the parameter space of the posterior probability to well-constrained values. As the November 2018 sample is the only observation falling into the MJD 58395-58492 window, a possible change in the observed rate across the active window is not a concern. Instead, we are exploring possible clustering on time scales of hours.\\

From the waiting times for the November 2018 sample shown in Figure~\ref{fig:pdf},  we see that the peak in $\delta t$ is roughly at $\sim$200 seconds. This set of 24 bursts during a single continuous 7-hr observation (see Table~\ref{tab:obs})  provides the most meaningful constraints out of the 13 observations from the Effelsberg dataset. 

Something peculiar about this observation is that most of the bursts were detected in the second half of the observation. If the 7-hour session is split into two sections of 3.5 hours each, the inferred Poisson rates are 27$\pm$5 day$^{-1}$ and 137$\pm$12 day$^{-1}$ for the first and second half, respectively.\\

We explore first the possibility of an observational bias leading to the observed disparity. Possible reasons are a change in the sensitivity of the telescope during the session, with the sensitivity improving in the second half and more RFI during the first half. We note that the observation started at UTC 19:58:50 at nearly $30^{\circ}$ in elevation, passed the zenith, and finished at $65^{\circ}$ at UTC 02:58:50 on the next day.\\

As the observations were done at 1.4 GHz, there is no gain-elevation effect due to the deformation of the antenna. However, the system temperature, $T_{\rm{sys}}$, increases by roughly 5\,K due to atmospheric effects (opacity $\tau=0.01$) at an elevation of $30^{\circ}$. The system temperature measured at the zenith is 21\,K. This change in $T_{\rm{sys}}$ leads to a decrease in the system equivalent flux density (SEFD) from 19.0\,Jy at zenith to 15.4\,Jy at $30^{\circ}$. This implies that a burst detected at zenith with an SNR of 8.5 could be easily missed as it would fall below detectability (SNR<7) at $30^{\circ}$ elevation. We neglect bursts with SNR < 8.5 from the second half of the observation (B13, B15, and B29) and obtain a new Poissonian rate of $116\pm$11 day$^{-1}$, which still does not match the event rate of the first part of the observation at the 3-$\sigma$ level.\\

Lastly, we check the influence of RFI. We inspect the mask files created by \texttt{rfifind} \citep{Ransom2011} for each one-hour long scan, and conclude that there is no obvious change in the RFI situation during the session. Furthermore, the number of candidates generated during the first and the second part of the observation, and find 8873 and 15074 respectively, which is in agreement with the expectation of more RFI at the higher elevations. A change in the RFI situation during the observation does not explain the higher rate inferred in the second part. From these checks, we conclude that the higher rate toward the end of the observation in November 2018 is likely not an observational bias.\\

We continue to explore the event rate asymmetry beyond Poissonian statistics, and in particular, if the detections of November 2018 are better fit by a Weibull distribution with $k<1$. As described in Section~\ref{subsec:weibull}, the best fit from the two dimensional posterior distribution for $k$ and $r$ predicts $k=0.62^{+0.10}_{-0.09}$, while if the one burst with $\delta t<1$\,s is excluded then $k'\,=\,0.73^{+0.12}_{-0.10}$. This strong dependence on the estimation of $k$ with few burst closely spaced is more evident with \citet{Gourdji2019} dataset,  where  the value shifts from $k\,=\,0.82^{+0.1}_{-0.09}$ to $k'\,=\,1.0^{+0.2}_{-0.1}$. In addition to the difference in the number of events included in the November 2018 and \citet{Gourdji2019} samples - 24 and 41 in total, respectively - the November observation was a single 7 hours long session, while the sample from \citet{Gourdji2019} comes from two sessions each of roughly 1.5 hours on consecutive days. Both observations are occurred at the center of their respective on-phase windows, with the November 2018 events at roughly $\phi=0.5$ (see Figure~\ref{fig:periodicity}) and \citet{Gourdji2019} at $\phi=0.55$. If we scale the Effelsberg event rate for November 2018 of 3.4 bursts/hr to AO's sensitivity we find a rate of 19 bursts/hr. The scaling considers the SEFDs described in Table~\ref{tab:technical} and the power low index of $\gamma=-1.1$ determined in Section~\ref{subsec: energy_distr}. The scaled rate is in fair agreement with the rate of 14 burst/hr from \citet{Gourdji2019}. Nonetheless, we cannot draw general conclusions on whether the center of the on-window leads to higher rates as we lack information on whether all the active windows have a similar activity behaviour.\\

Regarding the strong clustering reported by \citet{Oppermann2018} and recently by \citet{Oostrum2019}, we infer this was likely a consequence of the unknown periodicity and on-phase. If the analysis is limited to the active windows, despite some indication for small clustering, it is less obvious that it indeed differs from the Poissonian case. It is interesting that the exclusion of one wait time from the November 2018 dataset and two from the \citet{Gourdji2019} dataset reduces considerably the parameter space of posterior probability for $k$ and $r$ as seen in the top and bottom panels in Figure~\ref{fig:2Dposterior}, pre and post exclusion respectively. Particularly, the exclusion of two closely separated bursts from \citet{Gourdji2019} brings the distribution from a mildly clustered scenario to a distribution well described by Poisson statistics.\\ 

While excluding short waiting times allows us to investigate the change in $k$, their existence cannot be neglected. If the two distributions,  with $\delta t$<1\,s, and $\delta t$  of hundreds of seconds, are generated from different processes, the latter appears to be consistent with  Poissonian process. Another possible explanation, as hinted by B31, is that events separated by a couple of tens of milliseconds are in reality the two strongest components of broad bursts. If this is the case, considering the November 2018 and \citet{Gourdji2019} sample we conclude that the waiting time is still consistent with a Poisson distribution. Telescopes such FAST and Arecibo are crucial to discern this matter.\\

If on the contrary, the events with $\delta t$<1\,s are indeed independent bursts generated from the same mechanism as the ones separated by tens-to-hundreds of seconds, such mechanism needs to account for the high energy generation needed on a couple of milliseconds timescale. A high number of bursts on a given on-window is the ideal scenario to test how clustered the events are. To this end, telescopes such as Arecibo and FAST are also key given their sensitivity.\\

We stress that the combination of detections with different instruments can be misleading as the different sensitivities influence the event rate. An interesting study is to compare datasets of different telescopes over the same active window, or for a given telescope to compare the repetition pattern at different active phases to explore whether the event rate is constant or rather higher at given phases, such as the center of the active window. From the phase plot in Figure~\ref{fig:periodicity} we see that the number of events seems to be higher towards the center of the active phase. However, given the few epochs with detections, this estimation might not be significant.\\

Regarding how the energy of the bursts from FRB121102 is distributed, \citet{Law2017} calculated a power-law slope of $\gamma$\,$=$\,$-0.6^{+0.2}_{-0.3}$ for nine bursts detected by the Very Large Array at 3 GHz. The bursts energies ranged from $3\times10^{38}$\,erg to $9.8\times10^{39}$\,erg. In contrast,  \citet{Gourdji2019} came to a much steeper value of $\gamma$\,$=$\,$-1.8$\,$\pm$\,$0.3$ for a set of 41 bursts detected with Arecibo at 1.4 GHz. These bursts have an inferred energy ranging from $2\times 10^{37}$\,erg to $\sim$\,$2\times 10^{38}$\,erg, therefore probing a lower energy regime. Recently, \citet{Oostrum2019} came to a value of -1.7$\pm$0.6 from 30 bursts detected with WSRT/Apertif with energies in the range of $7\times10^{38}$\,erg to $6\times 10^{39}$\,erg.\\

\begin{figure*}
\begin{center}
\includegraphics[width=\textwidth]{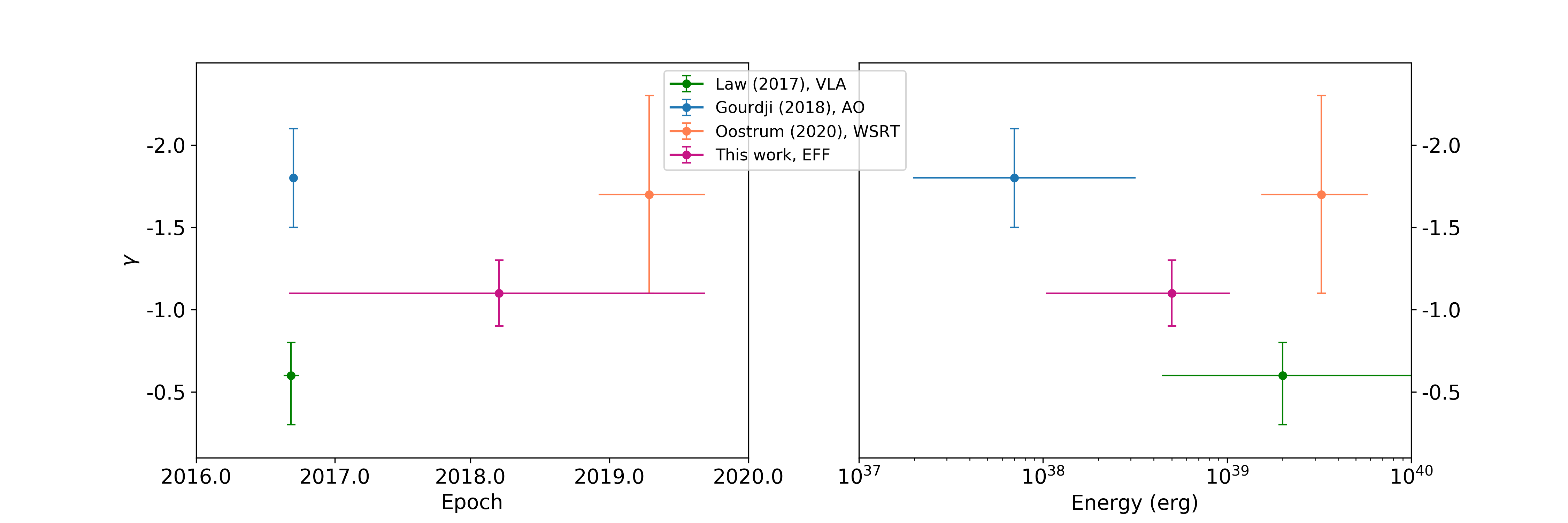}
\caption{Measured values for the slope of the power-law ($\gamma$) fit to the cumulative energy distribution of bursts from FRB 121102 as a function of time (left) and energy range (right)}. In green, the value is shown for the 9 bursts detected with the VLA at 3 GHz \citep{Law2017}, in blue for 41 bursts detected with AO at 1.4 GHz \citep{Gourdji2019}, in orange for 30 bursts from WSRT at 1.4 GHz \citep{Oostrum2019} and in magenta for the 57 bursts from Eff studied in this work. The error bars on the $\gamma$ values indicate 1-$\sigma$ uncertainties. The x-axis error bars indicate the time span of the dataset used to determine $\gamma$ (left) and energy span of the dataset used to determine $\gamma$ (right). 
\label{fig:gamma_values}
\end{center}
\end{figure*}

We show in Figure~\ref{fig:gamma_values} the  previously reported values for $\gamma$ in addition to our measurement (see Section~\ref{sec:bursts_prop}). The y-axis errors in both panels represent the 1-$\sigma$ uncertainties of $\gamma$. The x-axis bars denote the time span (left panel) and the energy range (right panel) of each dataset used for the gamma determination. Despite marginally agreeing if 3-$\sigma$ intervals are considered, converging to $\gamma\sim-1$, it is worth investigating the potential reasons for the different values encountered.\\

First, there is a strong dependence of the $\gamma$ value with the completeness limit used. Our measured value after rejecting the bursts below the completeness threshold and above the saturation limit is $\gamma=-1.1\pm0.2$. If the completeness threshold is not considered and the slope for all the bursts below the saturation limit is included, the slope flattens and becomes $\gamma=-0.8\pm0.1$. This is consistent with the $\gamma$\,$=$\,$-0.6^{+0.2}_{-0.3}$ reported by \citet{Law2017} and would misleadingly indicate a $\gamma$ near to the expected values for magnetars. \citet{Gourdji2019} and \citet{Oostrum2019} took into consideration such threshold for Arecibo and WSRT/Apertif respectively when reporting $\gamma$. It would be interesting to explore whether $\gamma $ changes for \citet{Law2017} dataset if the completeness limit is considered. Regarding a potential dependence on $\gamma$ with time, as proposed by \citet{Oostrum2019}, we see in Figure~\ref{fig:gamma_values} that the bursts considered for the Effelsberg dataset span roughly 3 years.  From Figure~\ref{fig:energy_dist} we see that all the bursts from the dataset are well mixed and follow the same trend, with no indication for different $\gamma$ with time.\\

The discrepancy of the values reported as of now challenge the universality of the power-law index for the cumulative energy distribution of the bursts and raises the question of whether we are in presence of a much more complex energy distribution (see right panel of Figure~\ref{fig:gamma_values}). Given that the energy range span of our dataset lies between the sample of low energy bursts of  \citet{Gourdji2019} and the more energetic bursts reported by  \citet{Law2017} and \citet{Oostrum2019}, one possibility is that a single power-law does not well describe the data over many orders of magnitude. In the right side of Figure~\ref{fig:gamma_values}, if we exclude the value reported by \citet{Law2017} - where the completeness threshold was not considered - we see that the slope of the energy distribution is steep for energies near $10^{40}\rm{erg}$ and $10^{38}\rm{erg}$, while being flatter in the intermediate energy range.\\

As the $\gamma$ value estimated from VLA comes from bursts detected at 3 GHz - contrary to the sample from AO, WSRT, and Eff at roughly 1.4 GHz - it could be that there is a dependency arising from the observing frequency. Additionally, the sensitivity of the instrument can play a role. Lastly, it can be that the observed energy distribution does not completely trace an intrinsic mechanism, but rather affected by propagation effects and observational biases. For instance, while \citet{Gourdji2019} and \citet{Oostrum2019} bursts were coherently de-dispersed, the \citealt{Law2017} and our Effelsberg sample are not. Perhaps not resolving the complex structure of some bursts leads to differences in the estimations of the widths of the bursts and therefore affecting their energy estimation.\\

We emphasize the importance of considering instrumental effects such as completeness and saturation limits, as well as the difference in sensitivities for the different telescopes when estimating the fit to the energy distribution of detected bursts.

\section{Conclusion}\label{sec:conclusion}
We have carried out an extensive 128-hour campaign with Effelsberg on the first repeater ever detected, FRB 121102, from September 2017 to June 2020. Some epochs are part of a multi-wavelength campaign to shadow telescopes at higher frequencies such as \nustar, INTEGRAL, and GTC. In total Effelsberg observed for 128 hr., Green Bank telescope 26 hr. and the Arecibo observatory 3.7-hr. At the time of the \nustar\ session on 2017-09-06 one burst was detected with Effelsberg: B6 (see Figure~\ref{fig:septemberbursts2017}) with flux of 1.56 $\pm$ 0.2 Jy\,ms. However, no X-ray photons were connected with such event. We placed upper limits on the energy of an X-ray burst counterpart to radio burst B6, which depends on the assumed spectral model (see Table~\ref{tab:nulims}). These limits are about an order of magnitude more constraining than those placed using {\em Chandra} and {\em XMM} by \citet{Scholz2017} for the case where X-ray emission is highly absorbed by material close to the source.\\

We combine our Effelsberg dataset with the published observations carried with identical setup by \citet{Hardy2017} and \citet{Houben2019}. The extended sample is composed of roughly 165 hours of observation from MJD 57635 to 59006 and consists of 34 epochs from which roughly 70\% are non-detections. Given that the observations were mostly randomly scheduled and that the full set of detections and non-detections are known, we searched for an underlying periodicity through the Lomb-Scargle periodogram. We find a 161$\pm$5 days period and an active window of roughly 54\%, which broadens to roughly 60\% after considering published observations at L-band. These values agree with the finding of \citet{PeriodicityFRB121102} on the potential 157 days periodicity for FRB 121102.\\

We continue to investigate how the time interval for consecutive bursts is distributed within the active windows, and particularly whether a Weibull distribution with $k<1$ (to allow clustering) describes the data better than the classic time-independent Poissonian statistics. To this end, we use the formalism implemented in \citet{Oppermann2018} over all the observations in the Effelsberg dataset lying within an active window. The November 2018 session is the only observation providing a reduced parameter space on the posterior probability for $k$ and $r$, mainly due to the high event rate and observation length. We observe a mild clustering in the sample, however, we find that the Weibull fitting is highly biased towards few events with waiting times shorter than one second. We test this finding with \citet{Gourdji2019} dataset which contains a total of 41 bursts over 3.2 hours of observation. We conclude that if the few shortly spaced events are removed, then $k=1$, meaning that the distribution is indistinguishable from a Poisson distribution. A bimodal distribution of the waiting time has been observed before by \citet{Li2019} and \citet{Gourdji2019} and could hint towards two mechanisms responsible for the events with $\delta t<1$\,s of separation and the ones that are hundreds of seconds apart.\\
An alternative scenario is that the events that are tens of milliseconds apart correspond to the main components of broad bursts in which weak, intermediate components are not detectable. This hypothesis is supported by burst B31 detected in November 2018, which with an FWHM of 39$\pm$2 ms is the widest burst ever measured for this source. We conclude that the strong clustering observed by \citet{Oppermann2018} and \citet{Oostrum2019} was a consequence of the unknown periodicity for FRB 121102.\\

Finally, we study the cumulative energy distribution of the 57 bursts from the Effelsberg dataset. We fit a power law of the form $N \propto E^\gamma$ through maximum likelihood analysis and find a slope of $\gamma=-1.1\pm0.1$. This value lies between the $\gamma$\,$=$\,$-0.6^{+0.2}_{-0.3}$ reported by \citet{Law2017} and $\gamma=-1.8$ from \citet{Gourdji2019}. Given the different energy regimes covered by the different studies, we suggest that a single power-law might not fit the data over many orders of magnitude or that the instrumental effects, such as completeness threshold and saturation, play an important role in its estimation. We find no indication for an epoch evolving  $\gamma$ as proposed by \citet{Oostrum2019}, as the bursts from the Effelsberg dataset are well mixed and described by a single power-law over the roughly 3 years of the data span.\\
\\
We finalize with key points to be considered after the results of this work:
\begin{enumerate}
    \item Given the existence of broad bursts with durations of tens of milliseconds, it is advised to future FRB searches to expand the parameter space of the burst widths to at least hundreds of milliseconds. For previously searched data, it is strongly encouraged to re-process the data.
    \item When doing a periodogram search it is important to compute the window transform to discard fake periodicities introduced, for instance, by the observation cadence (see Section~\ref{subsec:periodicity}) or noise-level. In addition to reporting the uncertainty of the measured periodicity (see Equation \ref{eq:sigma}) the false-alarm probability should be calculated as well. A conservative computational method is a bootstrap.
      \item We stress the importance of reporting non-detections in follow-up campaigns for any FRB. The knowledge of the start, duration, and outcome of the observation helps to better constrain the statistics for FRBs such as an underlying periodic active window and probability for detections of events.
    \item Instrumental effects such completeness limits, effects due to data conversion, and sensitivity should be taken into account when comparing energy distribution, event rates, waiting times, etc. across different telescopes.
\end{enumerate}
\section*{Acknowledgements}
We thank the referee for helpful and insightful comments that improved the quality of the manuscript.
We thank Dr. R. Wharton, Dr. N. Porayko, and L. Houben for helpful discussions. This publication is based on observations with the 100-m telescope of the Max-Planck-Institut für Radioastronomie at Effelsberg. We thank Dr. A. Kraus for scheduling our observations. This publication has received funding from the European Union’s Horizon 2020 research and innovation program under grant agreement No 730562 [RadioNet].
L.G.S. is a Lise Meitner independent research group leader and acknowledges funding from the Max Planck Society. J.W.T.H. acknowledges funding from an NWO Vici grant (``AstroFlash''). Partly based on observations with INTEGRAL, an ESA project with instruments and science data center funded by ESA member states. The Arecibo Observatory is a facility of the National Science Foundation operated under a cooperative agreement (\#AST-1744119) by the University of Central Florida in alliance with Universidad Ana G. Méndez (UAGM) and Yang Enterprises (YEI), Inc.


\section*{Data availability}
The data underlying this article will be shared on reasonable request to the corresponding author.



\bibliographystyle{mnras}
\bibliography{bibliography}

\begin{thebibliography}{}
\makeatletter
\relax
\def\mn@urlcharsother{\let\do\@makeother \do\$\do\&\do\#\do\^\do\_\do\%\do\~}
\def\mn@doi{\begingroup\mn@urlcharsother \@ifnextchar [ {\mn@doi@}
  {\mn@doi@[]}}
\def\mn@doi@[#1]#2{\def\@tempa{#1}\ifx\@tempa\@empty \href
  {http://dx.doi.org/#2} {doi:#2}\else \href {http://dx.doi.org/#2} {#1}\fi
  \endgroup}
\def\mn@eprint#1#2{\mn@eprint@#1:#2::\@nil}
\def\mn@eprint@arXiv#1{\href {http://arxiv.org/abs/#1} {{\tt arXiv:#1}}}
\def\mn@eprint@dblp#1{\href {http://dblp.uni-trier.de/rec/bibtex/#1.xml}
  {dblp:#1}}
\def\mn@eprint@#1:#2:#3:#4\@nil{\def\@tempa {#1}\def\@tempb {#2}\def\@tempc
  {#3}\ifx \@tempc \@empty \let \@tempc \@tempb \let \@tempb \@tempa \fi \ifx
  \@tempb \@empty \def\@tempb {arXiv}\fi \@ifundefined
  {mn@eprint@\@tempb}{\@tempb:\@tempc}{\expandafter \expandafter \csname
  mn@eprint@\@tempb\endcsname \expandafter{\@tempc}}}

\bibitem[\protect\citeauthoryear{{Bagchi}}{{Bagchi}}{2017}]{pulsar-asteriod}
{Bagchi} M.,  2017, \mn@doi [\apjl] {10.3847/2041-8213/aa65c9}, \href
  {https://ui.adsabs.harvard.edu/abs/2017ApJ...838L..16B} {838, L16}

\bibitem[\protect\citeauthoryear{Bannister et~al.,}{Bannister
  et~al.}{2019}]{Bannister2019}
Bannister K.~W.,  et~al., 2019, \mn@doi [Science] {10.1126/science.aaw5903}

\bibitem[\protect\citeauthoryear{{Barr} et~al.,}{{Barr}
  et~al.}{2013}]{Barr2013}
{Barr} E.~D.,  et~al., 2013, \mn@doi [\mnras] {10.1093/mnras/stt1440}, \href
  {https://ui.adsabs.harvard.edu/abs/2013MNRAS.435.2234B} {435, 2234}

\bibitem[\protect\citeauthoryear{{Bera} \& {Chengalur}}{{Bera} \&
  {Chengalur}}{2019}]{Bera2019}
{Bera} A.,  {Chengalur} J.~N.,  2019, \mn@doi [\mnras] {10.1093/mnrasl/slz140},
  \href {https://ui.adsabs.harvard.edu/abs/2019MNRAS.490L..12B} {490, L12}

\bibitem[\protect\citeauthoryear{{Bhandari} et~al.,}{{Bhandari}
  et~al.}{2020}]{askap2020}
{Bhandari} S.,  et~al., 2020, \mn@doi [\apjl] {10.3847/2041-8213/ab672e}, \href
  {https://ui.adsabs.harvard.edu/abs/2020ApJ...895L..37B} {895, L37}

\bibitem[\protect\citeauthoryear{{Burke-Spolaor} et~al.,}{{Burke-Spolaor}
  et~al.}{2012}]{Burke-Spolaor2012}
{Burke-Spolaor} S.,  et~al., 2012, \mn@doi [\mnras]
  {10.1111/j.1365-2966.2012.20998.x}, \href
  {https://ui.adsabs.harvard.edu/abs/2012MNRAS.423.1351B} {423, 1351}

\bibitem[\protect\citeauthoryear{{Caleb} et~al.,}{{Caleb}
  et~al.}{2020}]{Caleb2020}
{Caleb} M.,  et~al., 2020, arXiv e-prints, \href
  {https://ui.adsabs.harvard.edu/abs/2020arXiv200608662C} {p. arXiv:2006.08662}

\bibitem[\protect\citeauthoryear{{Chatterjee} et~al.,}{{Chatterjee}
  et~al.}{2017}]{Chatterjee2017}
{Chatterjee} S.,  et~al., 2017, \mn@doi [\nat] {10.1038/nature20797}, \href
  {http://adsabs.harvard.edu/abs/2017Natur.541...58C} {541, 58}

\bibitem[\protect\citeauthoryear{Chawla et~al.,}{Chawla
  et~al.}{2020}]{Chawla2020}
Chawla P.,  et~al., 2020, \mn@doi [The Astrophysical Journal]
  {10.3847/2041-8213/ab96bf}, 896, L41

\bibitem[\protect\citeauthoryear{{Cordes} \& {Wasserman}}{{Cordes} \&
  {Wasserman}}{2016}]{Cordes2016}
{Cordes} J.~M.,  {Wasserman} I.,  2016, \mn@doi [\mnras]
  {10.1093/mnras/stv2948}, \href
  {https://ui.adsabs.harvard.edu/abs/2016MNRAS.457..232C} {457, 232}

\bibitem[\protect\citeauthoryear{{Dhillon} et~al.,}{{Dhillon}
  et~al.}{2018}]{HiPERCAM2018}
{Dhillon} V.,  et~al., 2018, in \procspie. p. 107020L (\mn@eprint {arXiv}
  {1807.00557}), \mn@doi{10.1117/12.2312041}

\bibitem[\protect\citeauthoryear{{Dokuchaev} \& {Eroshenko}}{{Dokuchaev} \&
  {Eroshenko}}{2017}]{DNSmerger_repeat}
{Dokuchaev} V.~I.,  {Eroshenko} Y.~N.,  2017, arXiv e-prints, \href
  {https://ui.adsabs.harvard.edu/abs/2017arXiv170102492D} {p. arXiv:1701.02492}

\bibitem[\protect\citeauthoryear{{Gajjar} et~al.,}{{Gajjar}
  et~al.}{2018}]{Gajjar2018}
{Gajjar} V.,  et~al., 2018, \mn@doi [\apj] {10.3847/1538-4357/aad005}, \href
  {https://ui.adsabs.harvard.edu/abs/2018ApJ...863....2G} {863, 2}

\bibitem[\protect\citeauthoryear{{Gourdji}, {Michilli}, {Spitler}, {Hessels},
  {Seymour}, {Cordes}  \& {Chatterjee}}{{Gourdji} et~al.}{2019}]{Gourdji2019}
{Gourdji} K.,  {Michilli} D.,  {Spitler} L.~G.,  {Hessels} J.~W.~T.,  {Seymour}
  A.,  {Cordes} J.~M.,   {Chatterjee} S.,  2019, \mn@doi [\apjl]
  {10.3847/2041-8213/ab1f8a}, \href
  {https://ui.adsabs.harvard.edu/abs/2019ApJ...877L..19G} {877, L19}

\bibitem[\protect\citeauthoryear{{G{\"o}{\v{g}}{\"u}{\textcommabelow s}},
  {Woods}, {Kouveliotou}, {van Paradijs}, {Briggs}, {Duncan}  \&
  {Thompson}}{{G{\"o}{\v{g}}{\"u}{\textcommabelow s}} et~al.}{1999}]{Gogus1999}
{G{\"o}{\v{g}}{\"u}{\textcommabelow s}} E.,  {Woods} P.~M.,  {Kouveliotou} C.,
  {van Paradijs} J.,  {Briggs} M.~S.,  {Duncan} R.~C.,   {Thompson} C.,  1999,
  \mn@doi [\apjl] {10.1086/312380}, \href
  {https://ui.adsabs.harvard.edu/abs/1999ApJ...526L..93G} {526, L93}

\bibitem[\protect\citeauthoryear{{G{\"o}{\v{g}}{\"u}{\textcommabelow s}},
  {Woods}, {Kouveliotou}, {van Paradijs}, {Briggs}, {Duncan}  \&
  {Thompson}}{{G{\"o}{\v{g}}{\"u}{\textcommabelow s}} et~al.}{2000}]{Gogus2000}
{G{\"o}{\v{g}}{\"u}{\textcommabelow s}} E.,  {Woods} P.~M.,  {Kouveliotou} C.,
  {van Paradijs} J.,  {Briggs} M.~S.,  {Duncan} R.~C.,   {Thompson} C.,  2000,
  \mn@doi [\apjl] {10.1086/312583}, \href
  {https://ui.adsabs.harvard.edu/abs/2000ApJ...532L.121G} {532, L121}

\bibitem[\protect\citeauthoryear{{Hardy} et~al.,}{{Hardy}
  et~al.}{2017}]{Hardy2017}
{Hardy} L.~K.,  et~al., 2017, \mn@doi [\mnras] {10.1093/mnras/stx2153}, \href
  {https://ui.adsabs.harvard.edu/abs/2017MNRAS.472.2800H} {472, 2800}

\bibitem[\protect\citeauthoryear{{Harrison} et~al.,}{{Harrison}
  et~al.}{2013}]{hcc+13}
{Harrison} F.~A.,  et~al., 2013, \mn@doi [\apj] {10.1088/0004-637X/770/2/103},
  \href {https://ui.adsabs.harvard.edu/abs/2013ApJ...770..103H} {770, 103}

\bibitem[\protect\citeauthoryear{{Hessels} et~al.,}{{Hessels}
  et~al.}{2019}]{Hessels2019}
{Hessels} J.~W.~T.,  et~al., 2019, \mn@doi [\apjl] {10.3847/2041-8213/ab13ae},
  \href {https://ui.adsabs.harvard.edu/abs/2019ApJ...876L..23H} {876, L23}

\bibitem[\protect\citeauthoryear{{Houben}, {Spitler}, {ter Veen}, {Rachen},
  {Falcke}  \& {Kramer}}{{Houben} et~al.}{2019}]{Houben2019}
{Houben} L.~J.~M.,  {Spitler} L.~G.,  {ter Veen} S.,  {Rachen} J.~P.,  {Falcke}
  H.,   {Kramer} M.,  2019, \mn@doi [\aap] {10.1051/0004-6361/201833875}, \href
  {https://ui.adsabs.harvard.edu/abs/2019A&A...623A..42H} {623, A42}

\bibitem[\protect\citeauthoryear{{James}, {Ekers}, {Macquart}, {Bannister}  \&
  {Shannon}}{{James} et~al.}{2019}]{James2019}
{James} C.~W.,  {Ekers} R.~D.,  {Macquart} J.~P.,  {Bannister} K.~W.,
  {Shannon} R.~M.,  2019, \mn@doi [\mnras] {10.1093/mnras/sty3031}, \href
  {https://ui.adsabs.harvard.edu/abs/2019MNRAS.483.1342J} {483, 1342}

\bibitem[\protect\citeauthoryear{{Josephy} et~al.,}{{Josephy}
  et~al.}{2019}]{Josephy2019}
{Josephy} A.,  et~al., 2019, \mn@doi [\apjl] {10.3847/2041-8213/ab2c00}, \href
  {https://ui.adsabs.harvard.edu/abs/2019ApJ...882L..18J} {882, L18}

\bibitem[\protect\citeauthoryear{{Karuppusamy}, {Stappers}  \& {van
  Straten}}{{Karuppusamy} et~al.}{2010}]{Ramesh2010}
{Karuppusamy} R.,  {Stappers} B.~W.,   {van Straten} W.,  2010, \mn@doi [\aap]
  {10.1051/0004-6361/200913729}, \href
  {https://ui.adsabs.harvard.edu/abs/2010A&A...515A..36K} {515, A36}

\bibitem[\protect\citeauthoryear{{Keane}, {Stappers}, {Kramer}  \&
  {Lyne}}{{Keane} et~al.}{2012}]{giantpulses}
{Keane} E.~F.,  {Stappers} B.~W.,  {Kramer} M.,   {Lyne} A.~G.,  2012, \mn@doi
  [\mnras] {10.1111/j.1745-3933.2012.01306.x}, \href
  {https://ui.adsabs.harvard.edu/abs/2012MNRAS.425L..71K} {425, L71}

\bibitem[\protect\citeauthoryear{{Kraft}, {Burrows}  \& {Nousek}}{{Kraft}
  et~al.}{1991}]{kbn91}
{Kraft} R.~P.,  {Burrows} D.~N.,   {Nousek} J.~A.,  1991, \mn@doi [\apj]
  {10.1086/170124}, \href
  {https://ui.adsabs.harvard.edu/abs/1991ApJ...374..344K} {374, 344}

\bibitem[\protect\citeauthoryear{Kunjaya, Mahasena, Vierdayanti  \&
  Herlie}{Kunjaya et~al.}{2011}]{kunjaya2011}
Kunjaya C.,  Mahasena P.,  Vierdayanti K.,   Herlie S.,  2011, \mn@doi
  [Astrophysics and Space Science - ASTROPHYS SPACE SCI]
  {10.1007/s10509-011-0790-y}, 336

\bibitem[\protect\citeauthoryear{{Law} et~al.,}{{Law} et~al.}{2017}]{Law2017}
{Law} C.~J.,  et~al., 2017, \mn@doi [\apj] {10.3847/1538-4357/aa9700}, \href
  {http://adsabs.harvard.edu/abs/2017ApJ...850...76L} {850, 76}

\bibitem[\protect\citeauthoryear{{Lazarus}, {Karuppusamy}, {Graikou},
  {Caballero}, {Champion}, {Lee}, {Verbiest}  \& {Kramer}}{{Lazarus}
  et~al.}{2016}]{Lazarus2016}
{Lazarus} P.,  {Karuppusamy} R.,  {Graikou} E.,  {Caballero} R.~N.,  {Champion}
  D.~J.,  {Lee} K.~J.,  {Verbiest} J.~P.~W.,   {Kramer} M.,  2016, \mn@doi
  [\mnras] {10.1093/mnras/stw189}, \href
  {https://ui.adsabs.harvard.edu/abs/2016MNRAS.458..868L} {458, 868}

\bibitem[\protect\citeauthoryear{Li}{Li}{2010}]{RP_LI2010}
Li T.-H.,  2010, \mn@doi [Signal Processing]
  {https://doi.org/10.1016/j.sigpro.2010.01.012}, 90, 2133

\bibitem[\protect\citeauthoryear{{Li} et~al.,}{{Li} et~al.}{2015}]{Li2015}
{Li} Y.-P.,  et~al., 2015, \mn@doi [\apj] {10.1088/0004-637X/810/1/19}, \href
  {https://ui.adsabs.harvard.edu/abs/2015ApJ...810...19L} {810, 19}

\bibitem[\protect\citeauthoryear{{Li}, {Li}, {Zhang}, {Geng}, {Song}, {Huang}
  \& {Yang}}{{Li} et~al.}{2019}]{Li2019}
{Li} B.,  {Li} L.-B.,  {Zhang} Z.-B.,  {Geng} J.-J.,  {Song} L.-M.,  {Huang}
  Y.-F.,   {Yang} Y.-P.,  2019, arXiv e-prints, \href
  {https://ui.adsabs.harvard.edu/abs/2019arXiv190103484L} {p. arXiv:1901.03484}

\bibitem[\protect\citeauthoryear{{Liu}}{{Liu}}{2018}]{Liu2018}
{Liu} X.,  2018, \mn@doi [\apss] {10.1007/s10509-018-3462-3}, \href
  {https://ui.adsabs.harvard.edu/abs/2018Ap&SS.363..242L} {363, 242}

\bibitem[\protect\citeauthoryear{{Lorimer}, {Bailes}, {McLaughlin}, {Narkevic}
  \& {Crawford}}{{Lorimer} et~al.}{2007}]{Lorimer2007}
{Lorimer} D.~R.,  {Bailes} M.,  {McLaughlin} M.~A.,  {Narkevic} D.~J.,
  {Crawford} F.,  2007, \mn@doi [Science] {10.1126/science.1147532}, \href
  {https://ui.adsabs.harvard.edu/abs/2007Sci...318..777L} {318, 777}

\bibitem[\protect\citeauthoryear{{Macquart} et~al.,}{{Macquart}
  et~al.}{2020}]{Macquart2020}
{Macquart} J.~P.,  et~al., 2020, \mn@doi [\nat] {10.1038/s41586-020-2300-2},
  \href {https://ui.adsabs.harvard.edu/abs/2020Natur.581..391M} {581, 391}

\bibitem[\protect\citeauthoryear{{Marcote} et~al.,}{{Marcote}
  et~al.}{2017}]{Marcote2017}
{Marcote} B.,  et~al., 2017, \mn@doi [\apjl] {10.3847/2041-8213/834/2/L8},
  \href {http://adsabs.harvard.edu/abs/2017ApJ...834L...8M} {834, L8}

\bibitem[\protect\citeauthoryear{{Marcote} et~al.,}{{Marcote}
  et~al.}{2020}]{Chimeloc2020}
{Marcote} B.,  et~al., 2020, \mn@doi [\nat] {10.1038/s41586-019-1866-z}, \href
  {https://ui.adsabs.harvard.edu/abs/2020Natur.577..190M} {577, 190}

\bibitem[\protect\citeauthoryear{{Margalit}, {Berger}  \& {Metzger}}{{Margalit}
  et~al.}{2019}]{magnetar2019}
{Margalit} B.,  {Berger} E.,   {Metzger} B.~D.,  2019, \mn@doi [\apj]
  {10.3847/1538-4357/ab4c31}, \href
  {https://ui.adsabs.harvard.edu/abs/2019ApJ...886..110M} {886, 110}

\bibitem[\protect\citeauthoryear{{McLaughlin} et~al.,}{{McLaughlin}
  et~al.}{2009}]{McLaughlin2009}
{McLaughlin} M.~A.,  et~al., 2009, \mn@doi [\mnras]
  {10.1111/j.1365-2966.2009.15584.x}, \href
  {https://ui.adsabs.harvard.edu/abs/2009MNRAS.400.1431M} {400, 1431}

\bibitem[\protect\citeauthoryear{Mereghetti et~al.,}{Mereghetti
  et~al.}{2020}]{Mereghetti2020}
Mereghetti S.,  et~al., 2020, INTEGRAL discovery of a burst with associated
  radio emission from the magnetar SGR 1935+2154 (\mn@eprint {arXiv}
  {2005.06335})

\bibitem[\protect\citeauthoryear{{Metzger}, {Berger}  \& {Margalit}}{{Metzger}
  et~al.}{2017}]{mbm17}
{Metzger} B.~D.,  {Berger} E.,   {Margalit} B.,  2017, \mn@doi [\apj]
  {10.3847/1538-4357/aa633d}, \href
  {http://adsabs.harvard.edu/abs/2017ApJ...841...14M} {841, 14}

\bibitem[\protect\citeauthoryear{{Michilli} et~al.,}{{Michilli}
  et~al.}{2018}]{Michilli2018}
{Michilli} D.,  et~al., 2018, \mn@doi [\nat] {10.1038/nature25149}, \href
  {http://adsabs.harvard.edu/abs/2018Natur.553..182M} {553, 182}

\bibitem[\protect\citeauthoryear{{Morello}, {Barr}, {Stappers}, {Keane}  \&
  {Lyne}}{{Morello} et~al.}{2020}]{Morello2020}
{Morello} V.,  {Barr} E.~D.,  {Stappers} B.~W.,  {Keane} E.~F.,   {Lyne} A.~G.,
   2020, \mn@doi [\mnras] {10.1093/mnras/staa2291}, \href
  {https://ui.adsabs.harvard.edu/abs/2020MNRAS.497.4654M} {497, 4654}

\bibitem[\protect\citeauthoryear{{Oostrum} et~al.,}{{Oostrum}
  et~al.}{2020}]{Oostrum2019}
{Oostrum} L.~C.,  et~al., 2020, \mn@doi [\aap] {10.1051/0004-6361/201937422},
  \href {https://ui.adsabs.harvard.edu/abs/2020A&A...635A..61O} {635, A61}

\bibitem[\protect\citeauthoryear{{Oppermann}, {Yu}  \& {Pen}}{{Oppermann}
  et~al.}{2018}]{Oppermann2018}
{Oppermann} N.,  {Yu} H.-R.,   {Pen} U.-L.,  2018, \mn@doi [\mnras]
  {10.1093/mnras/sty004}, \href
  {https://ui.adsabs.harvard.edu/abs/2018MNRAS.475.5109O} {475, 5109}

\bibitem[\protect\citeauthoryear{{Petroff} et~al.,}{{Petroff}
  et~al.}{2016}]{FRBcat2016}
{Petroff} E.,  et~al., 2016, \mn@doi [\pasa] {10.1017/pasa.2016.35}, \href
  {http://adsabs.harvard.edu/abs/2016PASA...33...45P} {33, e045}

\bibitem[\protect\citeauthoryear{{Pilia} et~al.,}{{Pilia}
  et~al.}{2020}]{Pilia2020}
{Pilia} M.,  et~al., 2020, arXiv e-prints, \href
  {https://ui.adsabs.harvard.edu/abs/2020arXiv200312748P} {p. arXiv:2003.12748}

\bibitem[\protect\citeauthoryear{{Popov} \& {Stappers}}{{Popov} \&
  {Stappers}}{2007}]{Popov2007}
{Popov} M.~V.,  {Stappers} B.,  2007, \mn@doi [\aap]
  {10.1051/0004-6361:20066589}, \href
  {https://ui.adsabs.harvard.edu/abs/2007A&A...470.1003P} {470, 1003}

\bibitem[\protect\citeauthoryear{{Prochaska} et~al.,}{{Prochaska}
  et~al.}{2019}]{DSA2019}
{Prochaska} J.~X.,  et~al., 2019, \mn@doi [Science] {10.1126/science.aay0073},
  \href {https://ui.adsabs.harvard.edu/abs/2019Sci...366..231P} {366, 231}

\bibitem[\protect\citeauthoryear{{Rajwade} et~al.,}{{Rajwade}
  et~al.}{2020}]{PeriodicityFRB121102}
{Rajwade} K.~M.,  et~al., 2020, \mn@doi [\mnras] {10.1093/mnras/staa1237},
  \href {https://ui.adsabs.harvard.edu/abs/2020MNRAS.495.3551R} {495, 3551}

\bibitem[\protect\citeauthoryear{{Ransom}}{{Ransom}}{2011}]{Ransom2011}
{Ransom} S.,  2011, {PRESTO: PulsaR Exploration and Search TOolkit} (\mn@eprint
  {ascl} {1107.017})

\bibitem[\protect\citeauthoryear{{Ransom}, {Demorest}, {Ford}, {McCullough},
  {Ray}, {DuPlain}  \& {Brandt}}{{Ransom} et~al.}{2009}]{GUPPI2009}
{Ransom} S.~M.,  {Demorest} P.,  {Ford} J.,  {McCullough} R.,  {Ray} J.,
  {DuPlain} R.,   {Brandt} P.,  2009, in American Astronomical Society Meeting
  Abstracts \#214. p. 605.08

\bibitem[\protect\citeauthoryear{{Ravi} et~al.,}{{Ravi}
  et~al.}{2019}]{Ravi2019}
{Ravi} V.,  et~al., 2019, \mn@doi [\nat] {10.1038/s41586-019-1389-7}, \href
  {https://ui.adsabs.harvard.edu/abs/2019Natur.572..352R} {572, 352}

\bibitem[\protect\citeauthoryear{{Scholz} et~al.,}{{Scholz}
  et~al.}{2016}]{Scholz2016}
{Scholz} P.,  et~al., 2016, \mn@doi [\apj] {10.3847/1538-4357/833/2/177}, \href
  {https://ui.adsabs.harvard.edu/abs/2016ApJ...833..177S} {833, 177}

\bibitem[\protect\citeauthoryear{{Scholz} et~al.,}{{Scholz}
  et~al.}{2017}]{Scholz2017}
{Scholz} P.,  et~al., 2017, \mn@doi [\apj] {10.3847/1538-4357/aa8456}, \href
  {https://ui.adsabs.harvard.edu/abs/2017ApJ...846...80S} {846, 80}

\bibitem[\protect\citeauthoryear{{Spitler} et~al.,}{{Spitler}
  et~al.}{2014}]{Spitler2014}
{Spitler} L.~G.,  et~al., 2014, \mn@doi [\apj] {10.1088/0004-637X/790/2/101},
  \href {https://ui.adsabs.harvard.edu/abs/2014ApJ...790..101S} {790, 101}

\bibitem[\protect\citeauthoryear{{Spitler} et~al.,}{{Spitler}
  et~al.}{2016}]{Spitler2016}
{Spitler} L.~G.,  et~al., 2016, \mn@doi [\nat] {10.1038/nature17168}, \href
  {http://adsabs.harvard.edu/abs/2016Natur.531..202S} {531, 202}

\bibitem[\protect\citeauthoryear{{Tendulkar} et~al.,}{{Tendulkar}
  et~al.}{2017}]{Tendulkar2017}
{Tendulkar} S.~P.,  et~al., 2017, \mn@doi [\apjl] {10.3847/2041-8213/834/2/L7},
  \href {http://adsabs.harvard.edu/abs/2017ApJ...834L...7T} {834, L7}

\bibitem[\protect\citeauthoryear{{The CHIME/FRB Collaboration} et~al.,}{{The
  CHIME/FRB Collaboration} et~al.}{2020a}]{Chimeperiodicity2020}
{The CHIME/FRB Collaboration} et~al., 2020a, arXiv e-prints, \href
  {https://ui.adsabs.harvard.edu/abs/2020arXiv200110275T} {p. arXiv:2001.10275}

\bibitem[\protect\citeauthoryear{{The CHIME/FRB Collaboration} et~al.,}{{The
  CHIME/FRB Collaboration} et~al.}{2020b}]{CHIME2020_SGR}
{The CHIME/FRB Collaboration} et~al., 2020b, arXiv e-prints, \href
  {https://ui.adsabs.harvard.edu/abs/2020arXiv200510324T} {p. arXiv:2005.10324}

\bibitem[\protect\citeauthoryear{{Totani}}{{Totani}}{2013}]{DNSmerger}
{Totani} T.,  2013, \mn@doi [\pasj] {10.1093/pasj/65.5.L12}, \href
  {https://ui.adsabs.harvard.edu/abs/2013PASJ...65L..12T} {65, L12}

\bibitem[\protect\citeauthoryear{{VanderPlas}}{{VanderPlas}}{2018}]{LS2018}
{VanderPlas} J.~T.,  2018, \mn@doi [\apjs] {10.3847/1538-4365/aab766}, \href
  {https://ui.adsabs.harvard.edu/abs/2018ApJS..236...16V} {236, 16}

\bibitem[\protect\citeauthoryear{{van Straten} \& {Bailes}}{{van Straten} \&
  {Bailes}}{2011}]{DSPSR2011}
{van Straten} W.,  {Bailes} M.,  2011, \mn@doi [\pasa] {10.1071/AS10021}, \href
  {https://ui.adsabs.harvard.edu/abs/2011PASA...28....1V} {28, 1}

\makeatother
\end{thebibliography}

\bsp	
\label{lastpage}
\end{document}